\documentclass[a4paper]{article} 
\usepackage[top=1in,bottom=1in,left=1.1in,right=1.1in,asymmetric]{geometry}
\usepackage[utf8]{inputenc}
\usepackage{fancyhdr}
\usepackage[breaklinks]{hyperref}
\usepackage{graphicx}
\usepackage[backend=biber,style=phys,citestyle=numeric,sorting=none]{biblatex}
\usepackage{parskip}
\usepackage{indentfirst}
\usepackage{multicol}
\usepackage{amssymb}
\usepackage{amsmath}
\usepackage{chemformula}
\usepackage{wrapfig}
\usepackage{multicol}
\usepackage{wasysym}
\usepackage{caption}
\usepackage{subcaption}
\usepackage{pbox}
\newcommand{\TitleText}{Viability of a Dyson Swarm as a Form of Dyson Sphere}
\begin{document}

\begin{titlepage}
        \centering
        \begin{figure}[ht]
            \centering
            \includegraphics[width=0.25\textwidth]{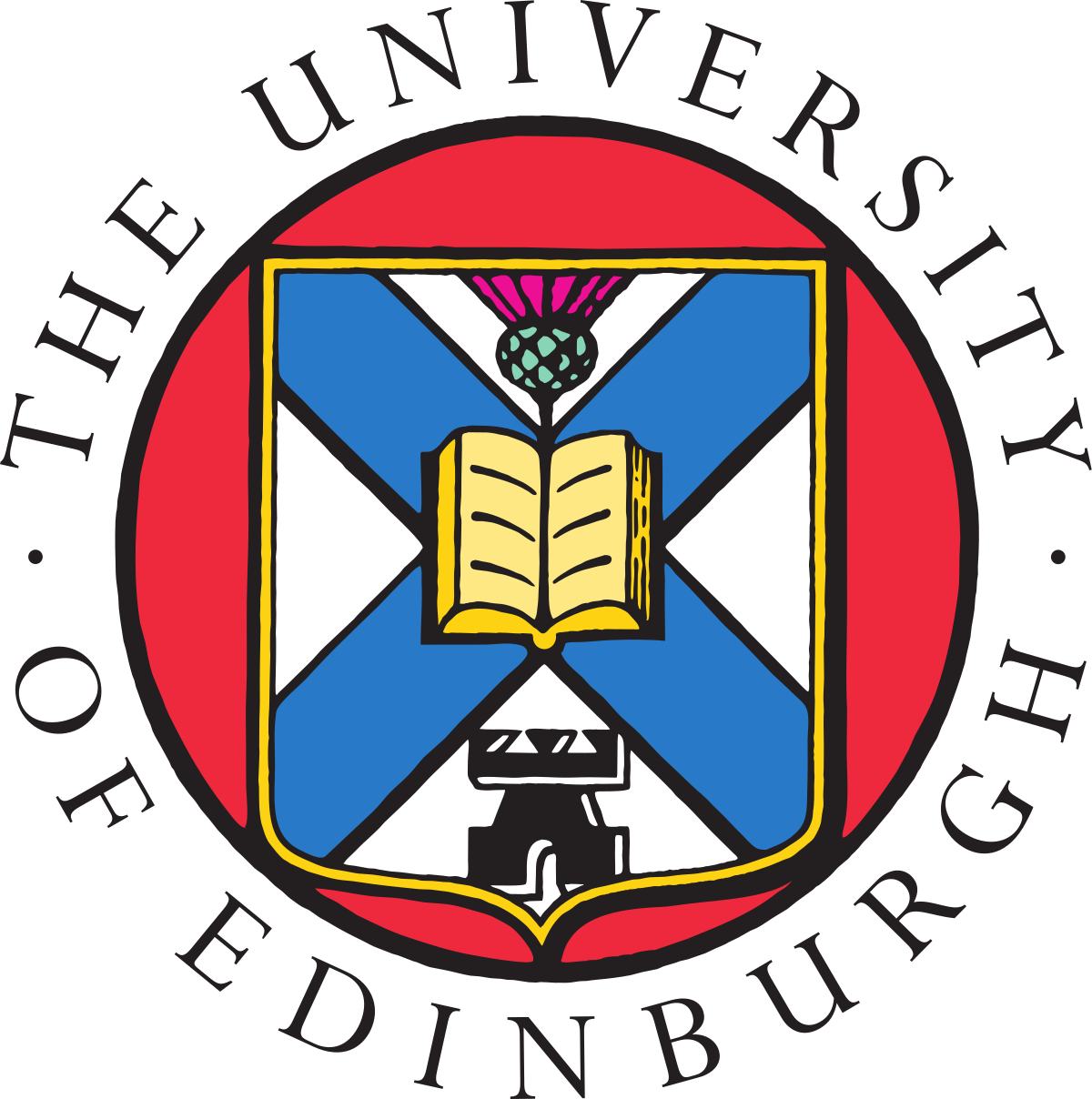}
        \end{figure}
    
        \vspace*{0.5cm}
    
        \huge\textbf{\TitleText}
    
        \vspace*{0.5cm}
    
        \large Jack Smith\footnote{Email: jackleesmith100@gmail.com}
        
        \textit{Undergraduate, School of Physics and Astronomy, University of Edinburgh}
        
        \today
        
        \hrulefill

        First conceptualised in Olaf Stapledon’s 1937 novel `Star Maker’, before being popularised by Freeman Dyson in the 1960s, Dyson Spheres are structures which surround a civilisation’s sun to collect all the energy it radiates. Through discussion of the features of such a feat of engineering, the viability, scale and likely design of a Dyson structure is evaluated, before details about each stage of its construction and operation are investigated. It is found that a Dyson Swarm, a large array of individual satellites orbiting another planetary body, is the ideal design for such a structure over the solid sun-surrounding structure which is typically associated with the Dyson Sphere.
        
        In our solar system, such a structure based around Mars would be able to account for the Earth’s 2019 global power consumption of 18.35 TW within fifty years once its construction has begun, which itself could start by 2040. The swarm of over 5.5 billion satellites would be constructed on the surface of Mars before being launched by electromagnetic accelerators into a Martian orbit. Efficiency of the Dyson Swarm ranges from around 0.74 - 2.77\% of the Sun's $3.85\times10^{26}$ W output, with large potential for growth as both current technologies improve and future concepts are brought to reality in the time before and during the swarm's construction. Not only would a Dyson Swarm provide a near infinite, renewable power source for Earth, it would allow for significant expansions in human space exploration and for our civilisation as a whole. 
        
        \section*{}
        
        \hrulefill
        
        \textit{Keywords: Dyson Sphere, Dyson Swarm, Wireless Power Transmission, Energy, Mars, Renewable Energy.}
        
        \vspace*{\fill}
    \end{titlepage}

\section{Introduction}
    
        \subsection{History and Design}
        
            Whilst outlining his ideas in the search for artificial sources of radiation in 1960, Freeman J. Dyson highlighted the energy and material requirements for an advanced civilisation, suggesting that should a planet’s material be used to surround its star, the solar radiation incident upon it could be harnessed and used. The `biosphere’ which Dyson described | with reference to our own Solar System | would be a hollow sphere constructed from Jupiter’s material, between 2 and 3 metres thick, surrounding the Sun and inner planets\cite{Dyson1960A}. Most of the Sun’s radiation would be incident upon this shell, providing the civilisation with a near-infinite source of energy. Indeed, it is the prime example of a type II civilisation on the Kardashev scale; one which obtains all of the energy radiated by its own star\cite{Kardashev1964}. Type I and III civilisations are those which are capable of collecting all of the energy from their own planet and galaxy, respectively. The Earth is not yet advanced enough for placement on this scale. Whilst this is the typical description of a Dyson Sphere, shortly after his publication Dyson clarified that “a solid shell or ring surrounding a star is mechanically impossible”, instead outlining how numerous smaller objects orbiting the star, could each collect small amounts of its radiation; this is now typically referred to as a `Dyson Swarm’\cite{Dyson1960B}.

            The largest issue with a single large shell is that of impacts with comets, asteroids, et cetera, which would break the structure, causing it to collapse into the Sun. The creation of a solid sphere would not only take an immense time, but the shell would also be liable to drift throughout the solar system as well as around and into the Sun, especially in its constructional stages of infancy. A swarm of smaller objects solves these issues as they can be individually produced much quicker and work independently of one another. They are also less vulnerable to impacts with natural solar satellites due to their size and numbers; the swarm would still function almost maximally should a few of its constituents be damaged or destroyed. A swarm would require less energy to produce and could be built to a smaller scale, whilst allowing for energy to be collected as the number of objects in the swarm is being increased, forming a positive feedback loop with parts of the energy being produced diverted back into the production of more so-called `swarm objects'.
        
        \subsection{Energy}
        
            A Dyson Sphere would be excessive to the amounts of energy we currently use due to the colossal energy output of the Sun. To estimate the size of a Dyson structure, we compare Earth’s annual energy usage to the Sun’s power output. In 2019, global energy consumption, $E_C$, was 583.90 exajoules (equivalent power consumption, $P_C = 18.5$ terawatts)\cite{BP2020}. The power output of the sun is found using the Stefan-Boltzmann law:
            
                \begin{equation}
                    P_\odot = A\sigma T^4 = 4\pi r^2\sigma T^4 = 3.85\times10^{26}\text{ W}
                \end{equation}
            
            where the surface temperature of the Sun is 5778 K\cite{Phillips1995}. The energy produced by the Sun in one second is hundreds of thousands of times greater than our species’ annual consumption. The main conclusion drawn from this is the inference that a Dyson Swarm is a much better and effective method than a complete Dyson Sphere. This is because using smaller individual objects will allow us to collect the quantities of energy we require without expending large amounts of resources and money into its construction. Should the sphere/swarm be positioned at an orbit of 1 AU (although this distance and location will be discussed later), the inverse square law dictates that the power received per unit surface area (intensity) at this distance will be:

                \begin{equation}
                    I_{\left(1\text{AU}\right)}= \frac{P_\odot}{4\pi r^2} = 1369\text{ W/m}^2
                \end{equation}
            
            To account for the Earth’s 2019 energy expenditure in the same time period, such a swarm (placed at 1AU from the Sun) would only have to surround:
            
                \begin{equation}
                    A=\frac{P_C}{ I_{\left(1\text{AU}\right)}}=\frac{18.5\text{ TW}}{1369 \text{ W/m}^2} = 1.35\times10^{10}\text{ m}^2
                    \label{eq:area_covered_at_earth}
                \end{equation}
            
            or $4.81\times 10^{-12}\%$ of the Sun; a large enough area, but the small percentage emphasises the unnecessary size of a complete Dyson Sphere. As Dyson himself suggested, a swarm is a much more possible structure than a sphere, which would not only be nigh on impossible to construct and maintain, but also far in excess of the ideal specifications of such a structure which we would consider producing. These estimates would change drastically depending on the details of the system used. For example, the distance of the swarm from the Sun would change the power each swarm object would collect and thus the number of swarm objects needed. Similarly, as with traditional solar panels, the objects are unlikely to be one hundred percent efficient. However, due to the immense scale of the Sun’s power output, their efficiency can be easily countervailed by increasing the number of objects. The adjustable size and scale of a Dyson Swarm, as well as its ease to produce and the reduced impact from contingencies, makes it the ideal form of a solar radiation-collecting structure.

\section{System Design}
    
    \subsection{Location}
    
        The main factor to be considered for the production and specification of a Dyson Swarm is its location, since this determines numerous factors about its design. The amount of material needed to produce the swarm puts the location on the planetary scale, where the swarm would be built on the surface from the material making up the mantle of that planet, before being launched into orbit at a similar radius to the planet around the Sun | though not necessarily a solar orbit at these radii. This close-to-planetary radius enables the minimum amount of energy possible to be expended in getting the swarm into orbit whilst also ensuring the simplicity of the swarm design, since significant movement in space would require a propulsion system to be built onto each object, complicating the construction process. Naturally, for a Dyson Swarm in our solar system, one of the eight planets would be utilised. The design of the structure is heavily dictated by the properties of the chosen planet, some of the most important of which are:
        
            \begin{itemize}
                \item surface gravity: this affects the amount of energy needed to launch the swarm into space
                \item orbital radius around the Sun: since the further away from the Sun the swarm objects are, the less radiation will be incident upon them, in turn meaning that more of them will need to be built
                \item composition: ideally, the materials used to construct the swarm objects will be obtained from the planet itself, thus the planet’s composition helps us determine the materials from which they may be made and thus whether the planet is suitable.
            \end{itemize}
    
        There are other, smaller factors such as surface temperature/conditions and the presence of an atmosphere, which may absorb amounts of incident radiation being reflected or increase the air resistance acting on the objects as they are launched, increasing the required energy. We can immediately rule out the four outermost planets of Jupiter, Saturn, Uranus, and Neptune. Not only do none of them have solid surfaces to land on, nor to build structures on, but their orbital radii about the Sun are too large. The closest of the four, Jupiter, has a mean distance from the Sun of 5.203 AU, giving a solar radiation intensity of:
    
            \begin{equation}
                I_{\jupiter}= \frac{P_\odot}{4\pi r^2} = 51\text{ W/m}^2
            \end{equation}
        
        which is twenty-seven times less than at Earth, meaning twenty-seven times as many objects would need to be produced, with a consequential increase in the amounts of energy and materials required. This decreases even further to $I_{\neptune}=1.51\text{ W/m}^2$ for Neptune (30.07 AU). The planetary nebula through which the planets were formed was mainly composed of hydrogen \& helium (98\%) with water, ammonia, methane accounting for the majority of the rest of its mass\cite{Taylor2009}. As a result, both the gas and ice giants (Jupiter \& Saturn, and Uranus \& Neptune, respectively) have very little metal or rock content, making them unsuitable for the location of the swarm. The four terrestrial planets are more suitable, each with advantages and disadvantages, as will be discussed below. Table~\ref{table:InnerPlanetsData} gives data on each of these four planets which are referred to continuously throughout sections \ref{section:EarthAndVenus} - ~\ref{section:Mars}.
    
            \renewcommand{\arraystretch}{1.5}
            \begin{table}[ht]
            \centering
            \caption{Planetary data for the four terrestrial planets. Data from
            \cite{NASA2019}, unless otherwise cited.}
            \label{table:InnerPlanetsData}
                \begin{tabular}{lcrrrr}
                \hline \hline
                 & & \textbf{Mercury, $\mercury$} & \textbf{Venus, $\venus$} & \textbf{Earth, $\earth$} & \textbf{Mars, $\mars$} \\ \hline
                Radius (km) & & 2440 & 6052 & 6378 & 3396 \\
                Surface Temp.\cite{Earthguide2017} (K) & [Max] & 722 & 738 & 331 & 293 \\
                & [Min] & 103 & 738 & 184 & 148 \\ \hline
                Surface Gravity (g) & & 0.377 & 0.905 & 1.000 & 0.379 \\
                Orbital Radius (AU) & & 0.390 & 0.720 & 1.000 & 1.880 \\
                Mass ($10^{24}$ kg) & & 0.330 & 4.870 & 5.970 & 0.642 \\
                Surface Pressure (bars) & & 0.000 & 92.000 & 1.000 & 0.010
                \end{tabular}
            \end{table}

        \subsubsection{Earth and Venus}
        \label{section:EarthAndVenus}
    
            Being the home of humanity, Earth is not the ideal location to produce the swarm. If the surface area of the swarm calculated in Eq.~\ref{eq:area_covered_at_earth} was built with one hundred percent efficiency to a depth of just one centimetre, the material required would take 2.9 m off the height of all land on Earth. At the same time, Earth has the highest surface gravity of the rocky planets, meaning that more energy will be required to lift the swarm objects into orbit, in turn increasing the size of the swarm to reach the desired energy output, which increases the amount of materials needed, and so forth. This effect is exacerbated by the Earth’s atmosphere, which makes it more energy expensive to reach escape velocity.
            
            Despite being 0.28 AU closer to the Sun and having a slightly reduced 91\% of the surface gravity of Earth, Venus is not much of a better alternative. The surface temperature on Venus is 738K, which rules out any human labour and will also limit the types of materials, machinery and infrastructure than can be used or built on the surface, since a melting point of well over 465 °C would be essential. A pressure of 92 bars is similarly problematic, demonstrated by the Venera 4 - 6 missions, which were all crushed by the atmospheric pressure before reaching the surface\cite{NASA2018}. The thick atmosphere also poses a problem since radiation which may be transmitted between the swarm, a base on Venus, and the Earth would be severely impeded as incident electromagnetic radiation to Venus would be mostly reflected back into space and likewise, radiation being sent out from the surface of Venus will be reflected back into the atmosphere. The compositions of both the Earth and Venus are made inconsequential by the severity of the planets’ aforementioned properties, which make them unsuitable options of a location to produce a Dyson Swarm.

                \begin{table}[ht]
                    \centering
                    \caption{Elemental composition of Mercury and Mars by percentage by weight (wt\%).}
                    \label{table:MercuryMarsCrustCompositions}
                        \begin{tabular}{crr}
                        \hline \hline
                        \textbf{Element} & \textbf{Mercury\cite{Solomon2018}} & \textbf{Mars\cite{Taylor2009}} \\ \hline
                        \ch{O} & 40.96 & 42.75 \\
                        \ch{Na} & 4.29 & 2.20 \\
                        \ch{Mg} & 10.00 & 5.46 \\
                        \ch{Al} & 6.92 & 5.56 \\
                        \ch{Si} & 29.26 & 23.00 \\
                        \ch{S} & 2.09 & $<$1 \\
                        \ch{Ca} & 4.39 & 4.95 \\
                        \ch{Fe} & 1.04 & 14.10
                        \end{tabular}
                \end{table}
    
    \subsubsection{Mercury}
    \label{section:Mercury}
    
        Being the closest planet to the Sun, Mercury receives the highest intensity of solar radiation of the eight planets at $I_{\mercury}=9003\text{ W/m}^2$, although a drawback of this is the high temperatures and amounts of radiation on its surface. These temperatures are exacerbated by Mercury’s long rotational period (59 days), compared to its orbital period about the Sun (88 days), which makes the length of one Martian day equivalent to 176 Earth days\cite{ESO2007}. Consequentially, each side of the planet reaches 449 °C and -170 °C during a Mercurian daytime and nighttime, respectively. These conditions are not only unsuitable for human habitation, but machinery and electronics would have to be built to withstand the intense heat and cold.
        
        The planet’s position and motion within the solar system also presents difficulties for reaching it. Whilst direct launch windows from Earth to Mercury occur approximately three times a year, Mercury’s high rotational speed and low gravitational field strength results in a high $\Delta V$ being needed for Mercurial orbit insertion\cite{Biesbroek2016}. This increases the amount of fuel needed for rockets, in turn raising the overall mass and cost. The lack of an atmosphere around Mercury makes full propulsion landings likely (as opposed to those which may be aided by a prior reduction in speed through account of drag in the atmosphere), further increasing the complexity and size of the rockets needed, which is a significant disadvantage due to the large quantities of machinery and infrastructure that would be sent to the planet to begin and support construction of the swarm. 
        
        Possibly the most important characteristics of Mercury and Mars (our two remaining candidates) are their compositions, as shown in Table~\ref{table:MercuryMarsCrustCompositions}. Though Mercury is referred to as the `Iron Planet’ by Sprague and Strom for having a planetary composition of 70\% iron and 30\% rocky material, the majority of this iron is not in the crust, but the core\cite{Strom2003}. The crust is between approximately 35 $\pm$ 10 km thick, making up $\sim$1.43\% of the planet’s radius\cite{Beuthe2020}. Iron makes up just 1\% of the crust on Mercury by weight, whereas silicon constitutes almost 30\% and is one of the main components of traditional solar cells. The seven and ten percent presence of aluminium and magnesium, respectively, are the main forms of metal on the surface. Aluminium and its alloys are commonly used for electronic components due to their high electrical conductivity as well as in construction\cite{RUSAL}. Launching the swarm into orbit from the surface of Mercury is made easy by the negligible atmosphere and its reduced mass compared to the Earth, which produces a 0.38g surface gravity, reducing the amount of energy that would be required to put the swarm into operation.
    
    \subsubsection{Mars}
    \label{section:Mars}
    
        Mars is the outermost of the terrestrial planets from the Sun which leads to its lower intensity of solar radiation of $I_{\mars}=387\text{ W/m}^2$. This distance, however, produces more favourable surface temperatures which range from 20 to -125 °C and are much more suitable for both humans and electrical systems. The very thin Martian atmosphere provides little protection from radiation and leaves the surface vulnerable to the effects of solar flares, with dust storms also traversing the planet. Though, as demonstrated by the numerous rovers and probes which have operated on Mars, these conditions do not limit the capabilities of the electrical equipment to a large degree. For example: dust accumulated on surfaces can be accounted for and occasionally blown off, and electronics are kept in a Warm Electronics Box to maintain operational temperatures\cite{NASARovers}. Human colonisation of other solar bodies is most synonymous with Mars and many of the challenges described above have various solutions, including living underneath the martial surface to reduce or block radiation and to provide insulation. Therefore, the surface conditions would allow for human labour on the surface and with missions planning to land humans on Mars by the end of the decade, the technology for such habitation is already being developed. Similarly, having sent numerous spacecraft to Mars, it is a well-understood process and much more is known about getting to Mars, and the planet itself, than that of Mercury.
        
        Launch windows to Mars open roughly every 2.1 years, meaning that large quantities of materials and machinery would need to be sent initially to provide means of constructing the basic infrastructure and getting started with construction of the swarm\cite{Biesbroek2016}. This first launch window would likely be completely autonomous and may include the construction of accommodation and facilities for humans to arrive to in the second launch window. Silicon and iron (23 wt\%, 14.1 wt\%) are the two most abundant elements after oxygen in the crust, which is around 50 $\pm$ 12 kilometres thick on average\cite{Taylor2009}. Both magnesium and aluminium have similar presences in the crust at about 5.5\% by weight. Launching the swarm is aided by Mars’ reduced gravitational field strength of 0.379g (almost identical to that of Mercury) though the thin atmosphere will increase drag on objects as they pass through, however, this is ideal for landing.

    \subsubsection{Comparison}
    \label{section:locationcomparison}
    
        With Mars and Mercury the two candidates, we can compare them through three categories: getting to them, building the swarm, and launching the swarm. Firstly, launching the swarm. Both planets have almost identical gravitational field strengths and though Mars’ atmosphere increases the energy needed to reach orbital velocities, the effect is not large enough to cause concern. One of the largest differences is the temperature, as the high temperatures from the Sun on the objects in orbit would affect the materials from which they could be made and the ways in which they would operate\cite{ESAMercury}. Such high temperatures are not an issue in orbit at Mars’ orbital radius, meaning the design of the swarm is less restricted to the materials it must use to combat the temperatures it will be vulnerable to. Thus, whilst both planets can be considered very similar in their ability to get the swarm into orbit, higher temperatures closer to the Sun give Mars the edge over Mercury as the design of the swarm objects is less confined, making the swarm easier to mass produce.
        
        In terms of getting to both planets, Mars is a much easier and efficient process. As mentioned above, Mercury’s orbital velocity around the Sun is the largest of the planets in the solar system, which makes it difficult to get into an orbit around it due to the 12.58 km/s $\Delta V$ required\cite{Biesbroek2016}. This would have to be accounted for via large amounts of fuel which increases mass dramatically, or by using other planets’ gravity to adjust course and speed, though this would take a long time and would depend on the positions of the planets. Mars on the other hand orbits slower than both Mercury and the Earth, however, so the amount of fuel and energy needed by rockets to get to Mars is significantly less with $\Delta V$ = 2.83 km/s for orbital insertion, making the transfer of materials from Earth easier. Similarly, the complete dependence on propulsion landings for craft arriving at Mercury further increases the amount of fuel and mass which would need to be accounted for, whereas this amount is lower on Mars as the air resistance of the atmosphere can be utilised to reduce the dependence on propulsion landings, which is highly beneficial. Whilst Mercury launch windows are six times more common than those to Mars, this encourages self-sufficiency from the Martian surface in the two-year periods between launch windows at which large quantities of machinery and materials would be sent to support phases of construction. In both cases, rockets such as SpaceX’s under-construction Starship will be one hundred percent reusable, helping to reduce costs and vitally keep space debris to a minimum in this large-scale operation, though launches of such a rocket to the innermost planet would have reduced payload capacities to account for the extra fuel needed. Overall, Mars is the much better option due to the easier process | which has already been done countless times by various spacecraft and missions | which in turn reduces the amount of fuel (allowing for larger payloads) and cost. Mercury can be kept as a secondary option, especially one which may be reconsidered at the time the swarm is being designed to account for advancements in technology, which may overcome one or both of the Mercurial temperatures and launch/landing challenges. 
        
        The construction and design of the swarm is the most significant factor when choosing a location, with the design of the swarm objects dependant on the composition of the planet’s crust and the efficiency, and thus size, of the swarm depending on the distance of the swarm from the sun. For example, a single swarm object at Mars’ mean orbital radius would need to have an area just over 23 times larger to collect the same amount of energy as one at Mercury’s respective distance. This increases either the size of each object or the total number of objects in the swarm, both of which increase the mass of materials needed, not to mention the extra energy and infrastructure that would be necessary to obtain this material on Mars, as compared to Mercury. However, there is a balance to be stuck between this increased energy demand on Mars to account for the reduced intensity of solar radiation, and the energy that would be consumed in reaching Mercury due to the difficulties described previously in reaching and landing on it. To see the effects of Mercury’s challenging orbital mechanics, we compare the costs and energy expenditures of sending masses to both planets. NASA’s Mars Reconnaissance Orbiter made up 0.31\% (by weight) of the Atlas V rocket which launched it, with its journey to Mars orbit lasting 7 months. On the other hand, the European Space Agency and Japan Aerospace Exploration Agency’s joint BepiColombo mission launched in October 2018 and is planned to reach orbit around Mercury in December 2025, a seven-year trip which includes nine flybys of the three innermost planets. The mission’s payload makes up 0.35\% of the Ariane 5’s launch mass, aboard which it was launched. Evidently, the BepiColombo mission is much more complex \& dependant on the positions of three planets; the vastly shorter travel time to Mars more than makes up for its two-year wait between launch windows. Whilst the masses of the payloads compared to the rockets’ launch masses are similar, neither account for methods of landing on the surface due to the nature of these missions, however, due to Mars’ atmosphere we know that the mass of fuel needed for Mercury landers will be higher, thus tipping the balance in Mars' favour. This also assumes that a gravity assist is possible every time rockets are launched to Mercury, if not, the fuel required would be much higher to account for the $\Delta V$ stated earlier, which will be the likely outcome since waiting for gravity assists will take even more time. Overall, Mars is again the better option for the journey and sending cargo due to the higher efficiency of rockets with respect to the mass of their payloads and launch masses. Such rockets will also reach Mars significantly quicker than Mercury, which is preferable, especially since any technologies sent to Mercury could become outdated before their arrival. As a sidenote, future rockets will be even more efficient and completely re-useable which will further increase efficiencies in fuel, mass and cost; SpaceX’s Starship and Super Heavy combination will be able to transport over one hundred tonnes of cargo to Mars’ surface\cite{SpaceX2020}.
        
        For the design of the swarm itself, there are two main design types for its constituent objects. The first is traditional solar panels (photovoltaic cells), of which the most common variant is crystalline silicon (c-Si) cells. This variant is a good design since silicon is the second most abundant element on both planets, and conducting metals could complete the cells. The downside is that they are reasonably complex to manufacture and are also typically about 20\% efficient, which increases the number of panels needed by a factor of five\cite{OEESolar}. An alternative design is to use mirrors which would | from orbit | reflect incident solar radiation onto central collecting stations on the planet’s surface, which could be many types of concentrated solar power stations. These would be much easier to produce due to their simple design and their reflectivity can be maximised to almost 100\% by using high reflection (HR) coatings\cite{Yenisoy2019}. Similar anti-reflective coatings can be used for solar cells which increase the amount of light which is absorbed by the cells, though these are typically included in c-Si cells\cite{Smets2015}. In both cases, solar radiation is initially emitted from the Sun before meeting the swarm objects, where it is either collected (by solar cells) or reflected to a certain location (mirrors). From here, the radiation from either design must then be transmitted to the location where it is processed further; this is ultimately the Earth, though the energy may pass through the host planet of the swarm beforehand. Thus, throughout each of these transmissions of radiation, the inverse square law will have a further, though reduced, effect on the total amount of energy able to be collected. In any case, both planets would be suitable locations for solar cell production, since both are rich in silicon and conductive metals, whereas reflecting mirrors would be better suited towards Mars where there is a higher abundance of iron, magnesium, and aluminium as well as compounds which would be ideal for high and anti-reflective coatings. 
    
        Through account of its preferable surface conditions, ease of access, and the availability of materials which allows for numerous swarm designs to be considered, Mars is the better choice to locate the production of the Dyson Swarm. Dependant on the technology available at the time of construction, however, Mercury should not be ruled out completely. 
    
    \subsection{Design}
    
        With Mars selected as the host planet, the design of the objects in Dyson Swarm can now be discussed in more detail, with the collection and energy discussed in section~\ref{section:3}. As outlined previously, the two main designs for the swarm objects are photovoltaic cells and reflecting sheets, both of which we will now refer to simply as `satellites’ over `swarm objects'. Any other satellites being described which are not those in the swarm will be labelled appropriately. Both methods have advantages and disadvantages compared to one another, but we can also look into how the required properties of the objects affect which design we choose, namely across three categories: construction, deployment, and operation. The construction of the satellites should be easy and require as little machinery and energy as possible, as this decreases the amount of infrastructure needed to be transported to Mars and the efficiency of the system as a whole. They should ideally also be constructed of the materials which are most readily available in the planetary crust | making the process straight forward and as independent as possible (ideally completely independent) from materials on Earth. For deployment, the satellites must be durable enough to withstand forces of launch and air resistance during ascent, and they also need to be capable of reaching a stable and possibly predetermined orbit. Once in orbit, they are very unlikely to be attended to, so must be durable and reliable. They may need their own controls and methods of movement, for example, reflecting sheets would need to rotate such that the radiation reflected is incident upon the required target, or transmitters from solar cells may need to point in a specific direction. With these characteristics in mind, we can form a design for both propositions.
        
    \subsubsection{Photovoltaic Cells}
    \label{section:pvcells}
    
        c-Si solar cells are the most common form of photovoltaic cells, with monocrystalline and polycrystalline designs\cite{Smets2015}. Whilst the former are the more efficient design, the polycrystalline (sometimes denoted multicrystalline) cells are simpler to produce and use silicon more efficiently, but do not have high heat tolerances, which we would require. The production process for monocrystalline cells is rather complex; firstly silicon dioxide (quartzite, specifically) is heated and mixed with carbon to isolate the pure (metallurgical) silicon which is then put through numerous processes, each of which has alternative options, to eventually produce doped silicon wafers\cite{Smets2015}. These wafers must then be cleaned through use of acidic or alkaline substances and multiple heating steps before the electrical contacts on the top of the wafer and the metallic backing are added. These are typically aluminium or silver, the former of which is common in the Martian crust. However, whilst many of the processes are viable to an extent on Mars (i.e. heating, cleaning), some materials and elements are more difficult to produce and obtain. For example, the two different processes by which the metallurgical silicon can be made into monocrystalline silicon ingots both require an inert atmosphere, the most abundant of which in Mars’ 0.01 bar atmosphere is Argon (1.9\%), which whilst possible, adds another layer of difficulty\cite{NASA2019B}. An even more significant problem is the lack of carbon on Mars (other than the 95\% carbon dioxide in the atmosphere) due to the dominance of the sulfur cycle over the carbon cycle, which is needed to initially produce the metallurgical silicon. Similarly, alternatives may have to be found for boron and phosphorous, which are commonly used to make the p- and n-type silicon, respectively, since these elements are scarce on Mars\cite{Taylor2009}. As the cells would need to be durable in space for a long time | like those on the International Space Station | they would need to be covered by a thin layer of glass to provide protection from radiation, which will slightly decrease their efficiency\cite{Bushong2016}. The same silicon dioxide used in the production of the c-Si cells can be used to produce this glass. 
        
            \begin{wrapfigure}{r}{0.2\textwidth}
                \includegraphics[width=0.2\textwidth]{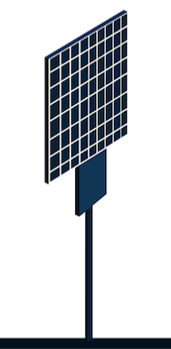}
                \caption{Render of one of Emrod's rectennae, held in the air by a support\cite{Emrod}.}
                \label{fig:rectenna}
            \end{wrapfigure}
        
        The energy absorbed by these solar panels will then need to be transmitted to Mars. This will likely be done using far-field radiative wireless power transfer (WPT) techniques, as they enable the losses from the dispersion of radiation to be reduced dramatically. Although they are not currently commonplace, we may assume that development of this technology would be included in the development of the Dyson Swarm. WPT would likely use microwaves as the carrier of energy; a transmitted microwave beam from the solar module in orbit would be received on the Martial surface\cite{Landis2006}. Two possible designs include a) Cassegrain antennae systems which send and receive radiation between the swarm and ground, or b) rectifying antennae (split into transmitting antennae and receiving antennae, the latter of which are often called rectennae) which consist of dipole antennae, such as those being introduced by Emrod\cite{Emrod,HU2021}. Both could be attached to the back of the solar modules although the latter, as shown in Fig.~\ref{fig:rectenna}, are flat which would be beneficial for keeping the profile of the satellites small for their launch to orbit. Whilst both of the WPT types discussed here are valid, the flat rectenna design will be adopted for use on the satellites themselves due to its more compact design and due to its current status and prospects in the real world. Large arrays of antennae can be formed, called antenna arrays, which allow for large transmitters/receivers whilst not being a solid structure and having the ability to be folded up and compacted for transportation. This WPT technology would also be used to transmit the power from the swarm back to Earth from Mars, for either a solar panel or reflecting panel swarm design and could use a mixture of antenna arrays and parabolic antennae depending on their location (the former for space-based applications, the latter for ground-based). Its use in this capacity will be discussed later.

    \subsubsection{Reflecting Sheets}
    \label{section:reflectingsheets}
    
        The design of the reflecting sheets as described above, is simply a large sheet of reflective material, for example iron, which may be coated with an anti-reflective coating to increase reflectivity. This sheet can be flat or shaped into a parabolic mirror to send the radiation incident upon it to collecting stations on the surface of the host planet. To optimise this reflectivity, however, we must know the details of the collecting stations which will convert the incident radiation into energy/electricity, as this determines which parts of the electromagnetic radiation spectrum we will be using, in turn changing the materials we may wish to use for the reflecting sheets. 
        
\begin{figure}[ht]
                \centering
                \includegraphics[width=0.6\textwidth]{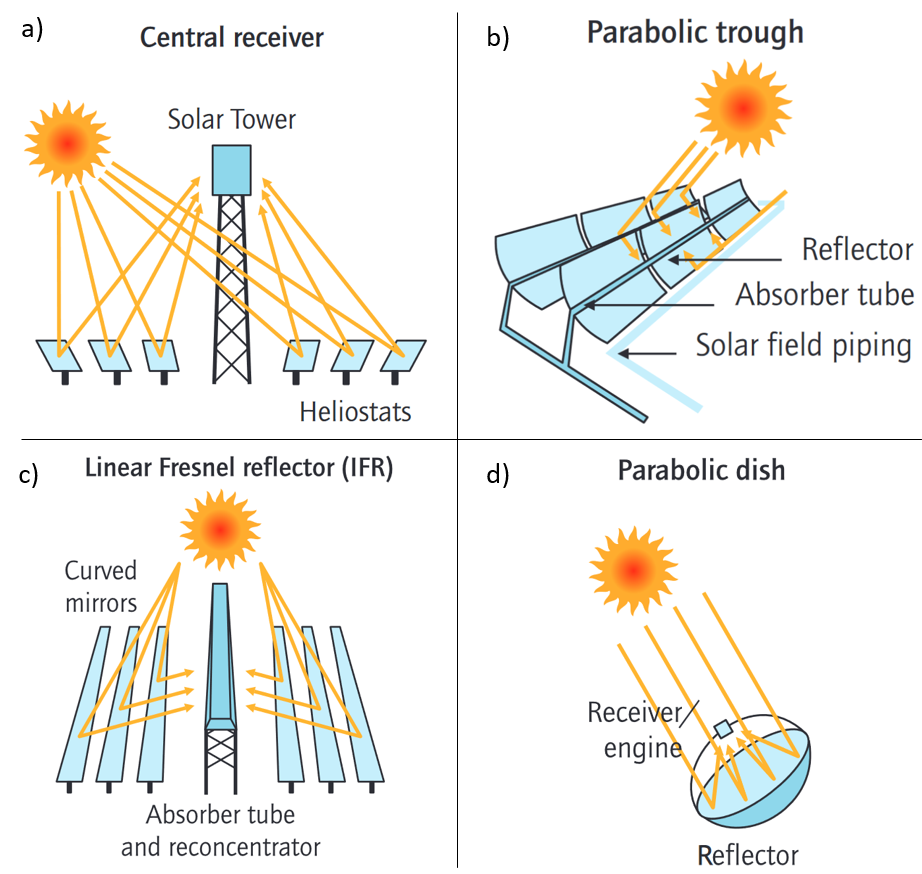}
                \caption{\cite{IEA2010}. a) A Solar Power Tower (SPT) consists of reflectors which focus the radiation onto a central receiver containing a heat transfer fluid (HTF) which is heated and used to power turbines (i.e. steam turbines). b) Parabolic Trough Collectors (PTC’s) reflect and focus radiation onto an absorber tube with a HTF inside, following the same mechanics as SPT’s. c) The Linear Fresnel Reflector (LFR) design again uses a linear absorber tube with a HTF, with radiation being incident upon it following reflection from rows of mirrors. d) With Parabolic Dish Collectors (PDC’s) each parabolic mirror focuses solar radiation to their own focal point at which a receiver containing an engine/generator with a working fluid such as a Stirling engine or a microturbine.}
                \label{fig:CSPTypes}
            \end{figure}

        Knowing that the solar radiation will be reflected towards the surface of Mars, we must converge upon a design for the collectors which will convert this incident radiation into useful energy. Concentrated solar power (CSP) systems such as those outlined by H. Zhang et al. in \cite{Zhang2013} provide various methods to realise this goal, some thermoelectric generation designs are illustrated with brief descriptions in Fig.~\ref{fig:CSPTypes}, though they are all based on the primary concept of using mirrors to reflect radiation onto receivers/engines. Often, CSP methods have built-in thermal storage capabilities which enable the short-term storage of thermal energy (e.g. via molten salts) which can then be used at a later time when the solar output alone is not sufficient\cite{IEA2010}. On Earth, this is used when the Sun sets and the stored energy can be used to continue the power output of the whole system despite the lack of sunlight, however, in the context of the Dyson Swarm such thermal storage systems would not be needed as there is no requirement for a continuous power output since the electricity produced will be transferred and/or stored before being sent back to Earth. 
        
        Solar cells, however, should not be ruled out for the CSP stations because as there are likely to only be a few of these stations, the solar panels used in them would not necessarily need to be produced on Mars but could instead be sent from Earth. Such solar cells could be laid in an array surrounded by more reflecting mirrors like the designs in Fig.~\ref{fig:CSPTypes}, though clearly the solar tower, receivers, etc. would be a collection of solar cells in this instance. This limits the number of solar cells which would be required, since a large array of mirrors would allow for the satellites to transmit to a large area on the ground whilst ensuring that the radiation will still end up incident on the solar cells. For an arbitrary example, if the full width at half maximum (FWHM) of the radiation reflected from a satellite in the swarm is, say 50 m, a 200 m radius array of mirrors could be laid out with solar cells at their focus. Such mirrors used on the ground would be easy to construct since they would be of the same design as those in orbit, just of a different shape and size to optimally focus light onto their target. 
        
        Both approaches (thermal \& solar) have their benefits and drawbacks. Stirling engines | used as a base for the thermal approaches | have been demonstrated to run for over fourteen years without maintenance and with no degradation\cite{NASA2020B}. After the same fourteen years, a PV module (with estimated rates of degradation of monocrystalline silicon PV modules being between 0.36 - 0.47\% per year\cite{Jordan2013}) can be estimated to have degraded by 4.92 - 6.38\%. The base efficiencies of the two electricity generation methods are similar, with Stirling engines just taking the edge at 21.4\% to PV cells’ 20.5\%\cite{AWAN2021}. The current best Stirling engines have efficiencies up to 32\%, with theoretical peak efficiencies reaching near to 40\% possible, which highlights the space into which this technology can improve in the time before the swarm is built\cite{BREEZE2018}. c-Si Solar cells have only been demonstrated to reach a maximum efficiency of 26.7\%, with a limit at around 30\%\cite{Andreani2019}, however, since both technologies would be implemented on the Martian surface only, advancements in their efficiencies can easily be implemented to the swarm by simply upgrading/replacing the outdated technology on the ground, the effect of which is received by the swarm as a whole. PV modules sent to Mars from Earth would be able to be set to use immediately and independently of any other infrastructure on the ground. This is applicable, since a much smaller number of solar cells would be required for ground use than if each satellite had solar cells built-in, in which case they would not be able to be sent from Earth; the smaller scale of solar cell use in this reflecting design allows for this possibility. One of the main problems of solar cells being covered by Martian dust can be combatted by human or autonomous cleaning. The Stirling engines could again be sent from Earth provided the number required is low, such that the transportation between planets would be more efficient than if machinery and infrastructure were to be utilised to produce the engines on the surface. At the same time, the working fluid within the engines is likely to be helium or hydrogen due to their abilities to absorb and release heat quickly, however, neither of these two elements make up a large proportion of the Martian atmosphere, further suggesting either the ease of production on Earth and transportation to Mars, or the benefit of solar cells\cite{BREEZE2018,HABERLE2015}. In principle, both are considerable designs and since the rest of the details about the Dyson Swarm are not reliant on this exact choice, no selection is made, but an efficiency of 21\% is adopted for the conversion of radiation to electricity. 
        
            \begin{wrapfigure}{r}{0.2\textwidth}
                \includegraphics[width=0.2\textwidth]{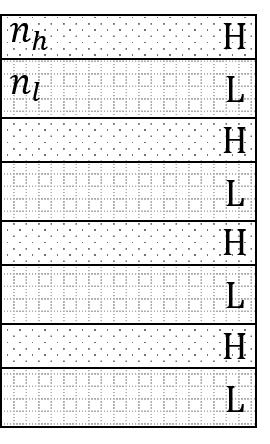}
                \caption{Diagram of a dielectric coating and its high and low refractive index layers.}
                \label{fig:hrc}
            \end{wrapfigure}
        
        With the design of the CSP aspect of the swarm laid out, the design of the reflecting sheets themselves can now be considered. Through the above discussion, we know that the radiation incident upon them needs to be directed onto a collection of mirrors (for either solar or thermal approaches). A parabolic mirror in orbit is a must since the incident radiation can then be focused directly to a point on the surface (bearing in mind the dispersion of this radiation), the precision of which flat mirrors are not able to achieve. A singular parabolic mirror facing the Sun would simply reflect the radiation back in that direction, so at least one smaller secondary mirror is required as well as a collimator or focusing lens which would then send radiation to the ground. The diameter of the mirrors will only depend on the limitations imposed on their size through their construction and launch processes, i.e. if they can fit within their launch vehicle, since there is no required amount of radiation they must reflect to function. A collimator would have to be built such that the divergence of the beam is minimal over the orbital distance it must cover to ensure the highest amount of radiation is incident on the CSP stations. This hardware would also require a method of movement to point towards the ground stations. The underlying principle of the mirrors is to reflect as much radiation as possible. To achieve this, high reflective coatings (also called dielectric coatings) can be added onto the surface of a substrate in order to increase their reflectivity to over 99.99999\%\cite{Prasad2018}. This is achieved by stacking layers of two materials with high and low refractive indices, as shown in Fig.~\ref{fig:hrc} where `H’ and `L’ signify the high and low refractive index materials, respectively. The system is called a quarter waveplate, as the individual layers are of physical thickness, $\delta$, which is a quarter of the wavelength of the incident EM radiation, $\lambda$:

            \begin{equation*}
                \delta=\frac{\lambda}{4n}
            \end{equation*}
        
        where $n$ is the refractive index of the layer. The quarter waveplate ensures that constructive interference (when phase difference between waves is $2n\pi$ where $n$ is an integer) occurs at the boundary between each pair of layers.
        
        \captionsetup{width=.4\textwidth}
            \begin{figure}[ht]
              \centering
             \begin{minipage}[t]{0.495\textwidth}
                 \centering
                 \includegraphics[width=\textwidth]{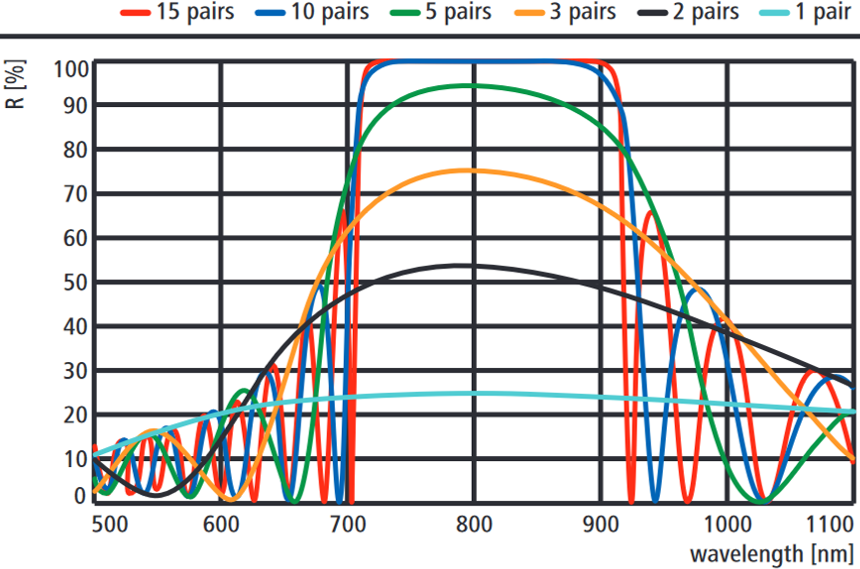}
                 \caption{Reflectance of a \ch{Ta2O5/SiO2} quarter wave stack at 800 nm, with increasing numbers of layers pairs (1 pair is one layer of each material)\cite{LAYERTEC}.}
                 \label{fig:dielectricgraph}
             \end{minipage}
             \hfill
             \begin{minipage}[t]{0.495\textwidth}
                 \centering
                 \includegraphics[width=\textwidth]{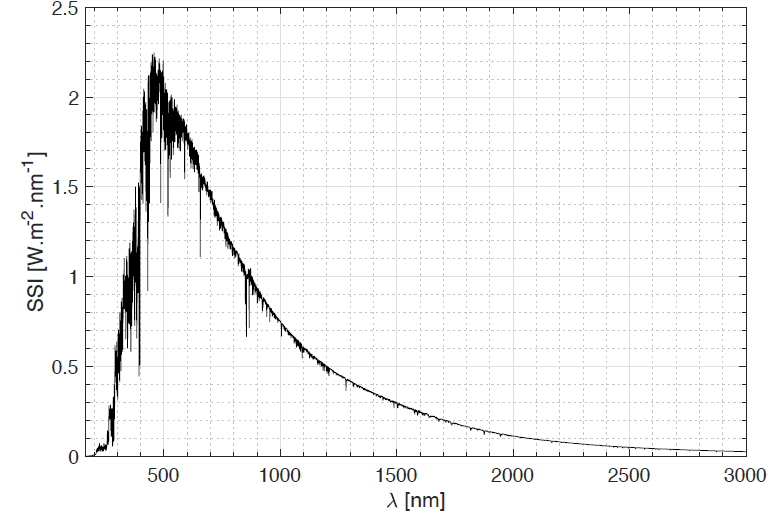}
                 \caption{Solar spectrum\cite{ESA2017}.}
                 \label{fig:solarspectrum}
             \end{minipage}
            \end{figure}
        \captionsetup{width=.8\textwidth}

        Increasing the number of pairs of layers increases the total reflectivity of the dielectric coating, as shown in Fig.~\ref{fig:dielectricgraph}. Since the thickness of the layers depends on the wavelength of the incident waves, this wavelength must be identified in order to suitably produce a dielectric coating of the required parameters. Since the method of electricity production is through the conversion of thermal energy, the wavelength of the EM spectrum used should be the one which transmits the most thermal energy (for a thermoelectric approach), which is infrared radiation, or one which is above the silicon band-gap energy wavelength of 1.1 µm (for solar c-Si cells). The IR part of the EM spectrum ranges from around 700 nm to 1 mm, though the drop-off in intensity of the solar spectrum after its visible light peak | as illustrated by Fig.~\ref{fig:solarspectrum} | indicates that a shorter infrared wavelength of a few micrometers would be ideal. Materials commonly used for IR dielectric coatings include \ch{ZnO }, \ch{ TiO2}, \ch{MgO} and \ch{Al2O3}, all but the former of which are present in the Martian crust\cite{Taylor2009, Prasad2018}. Table~\ref{tab:refractivematerials} shows some properties of these materials, as well as silicon dioxide, which is not only the most abundant material in Mars’ crust, but also has a lower refractive index than the other materials in discussion.

            \begin{table}[ht]
                \centering
                \caption{Refractive indices\cite{REFRACDATA} for given materials at specified wavelengths and their percentage by weight\cite{Taylor2009} of the Martian crust.}
                \label{tab:refractivematerials}
                \begin{tabular}{lcccr}
                \hline \hline
                \textbf{Material} & \multicolumn{3}{c}{\textbf{Refractive Index}} & \textbf{wt\%} \\ 
                 & 1.5 µm & 2 µm & 2.5 µm &  \\ \hline
                \ch{TiO2} & 2.2829 & 2.2647 & 2.2477 & 0.98 \\
                \ch{Al2O3} & 1.7470 & 1.7377 & 1.7262 & 10.50 \\
                \ch{MgO} & 1.7153 & 1.7085 & 1.7008 & 9.06 \\
                \ch{SiO2} & 1.4446 & 1.4381 & 1.4298 & 49.30
                \end{tabular}
            \end{table}

        Though there are numerous other materials which will be suitable for the dielectric mirrors, with these four materials being the most freely available on Mars, they are the best and easiest options to go for. Titanium dioxide and silicon dioxide are the best of the selection as they have not only the highest and lowest refractive indices respectively but are the most abundant in the crust. Both \ch{MgO} and \ch{Al2O3} could be used in place of the silicon dioxide, though the number of film layers in the dielectric coating may need to be increased to account for their higher refractive indices than the \ch{SiO2}. The substrate onto which the dielectric coating would be applied is commonly either glass or silicon wafers\cite{Prasad2018} although, as encountered with PV cells, silicon wafers are complex to produce. Glass could be used due to the abundance of silicon dioxide, though an easier option may be to use iron, aluminium or magnesium (or their alloys, such as ferroaluminium) since they have much lower melting temperatures than silicon dioxide and are not as fragile as glass. These metals also have high reflectivities by themselves: iron 65\%; magnesium 74\%; aluminium 70\%, whereas silicon wafers reflect around 55\% of 800 nm radiation incident upon them | a lower reflectivity for a more complicated process\cite{Ozdemir2011,Samsonov2012}.
        
        Thus, we can create a final design for the reflecting sheet route of the Dyson Swarm. The swarm would be made up of parabolic mirrors with a collimator/focuser in Mars orbit which focus solar radiation to a central collecting station on the Martian surface; a collection of mirrors which reflect radiation onto solar arrays or thermoelectric engines. The dishes would be constructed from one, or a mixture of: iron, aluminium, and magnesium with a dielectric coating of \ch{TiO2} and \ch{SiO2} to maximise reflectivity. The metallic substrate will be as thin as possible to ensure a minimal amount of material is used as possible, but thick enough to be rigid and support the equipment on it; a thin parabolic sheet with a triangular infill pattern may be suitable. The reflection of light is unlikely to be 100\% efficient unless the size of the target area on the ground is larger than the width of the reflected beams following their divergence between the satellites and the ground; this could be achieved by a large solar array, but is heavily dependant on the divergence of the beams. Since the individual satellites have no method of storing the radiation, they must always be in contact with a ground station to contribute to the swarm, though a method of stopping transmission from the satellites may be beneficial, especially when they move to target a new ground station or in special circumstances. This signifies that numerous stations would need to be constructed across the surface, with the satellites switching to the nearest as they orbit. The specific design of these satellites will be discussed further and finalised in section~\ref{section:3}.
        
    \subsubsection{Comparison}
    \label{section:comp2}
    
        There are some features which would be shared between both designs, such as small methods of performing attitude corrections and adjustments, like Hall thrusters, as well as the requirement for built-in batteries or other power systems to run any electric components and control systems. Common lithium-ion batteries are unlikely to be used due to the lack of lithium in the Martial crust, however, magnesium batteries with manganese dioxide cathodes could be a substitute due to the abundance of both \ch{Mg} and \ch{Mn} in the crust\cite{Taylor2009,Ling2017}. Such magnesium batteries have high energy densities and have a long lifespan, as required by this application. Both designs may require methods to radiate heat away through use of radiator panels as well as a mode of communication with the ground, possibly through orbital relay satellites. The dielectric coating used to enhance reflectivity on the parabolic mirrors will be similar to the anti-reflection coating used on the solar panels, thus there is no advantage over one or the other.
        
        A shared feature hitherto undiscussed is the need for control of the orientation of the objects in orbit such that solar cells face the optimum angle to the sun and so mirrors are continuously at their optimum working angle. In order to achieve this, reaction control wheels (RCW’s) are one of the best choices since they do not rely on fuel (gaseous/liquid). Both designs allow for the powering of such RCW’s and other electronic control systems through the onboard batteries. An issue with RCW’s is that they can become saturated with excess angular momentum, which needs to be cancelled out. This problem could be combatted by use of magnetorquers (devices which produce a torque through the interaction of a magnetic field with a magnetic momentum which they generate\cite{Tregouet2014}. Whilst Mars does not produce its own geomagnetic field, its induced magnetosphere as a result of solar winds could be used to provide such a torque for desaturation of the RCW’s, but the variability of this field, due to the continuous motion of the satellites around the planet, is a challenge for magnetorquers\cite{NASA2020,Tregouet2014}. RCW’s also add a mechanical element of redundancy to the design of the satellites, but this is not a large enough concern to switch to fuelled methods of attitude control such as cold gas thrusters, especially as the orientation of the satellites would continuously change, depleting fuel reserves rapidly. Other components of the satellites such as collimators and WPT transmission antennae would need to be electronically controlled in a separate but precise manner to point in any required direction. 
        
        Clearly, the design of the parabolic mirrors is much simpler, as well as the processes through which they would be made. The methods used to produce the c-Si cells are complex and require numerous materials and conditions, whereas the mass production of simple metallic sheets can be quickly achieved without significant difficulties. The other electronics | apart from the WPT of the solar cells | are shared between both designs, but this lack of WPT for the parabolic reflecting mirrors is their significant advantage as WPT introduces an inefficiency, in turn producing heat, whereas the dispersion of reflected radiation does not, not to mention its much more simplistic design. On the large scale of this project, the production process needs to be as easy as possible to allow for the highest efficiency in production and deployment of the swarm and to also reduce the amount of machinery which is required to be operational on the host planet. The less material we have to send ourselves, the less energy and money will be expended. This is the crucial advantage of the parabolic mirrors over PV modules, as the latter are much more complicated to produce.  
        
            \begin{figure}[h]
                \centering
                \includegraphics[width=\textwidth]{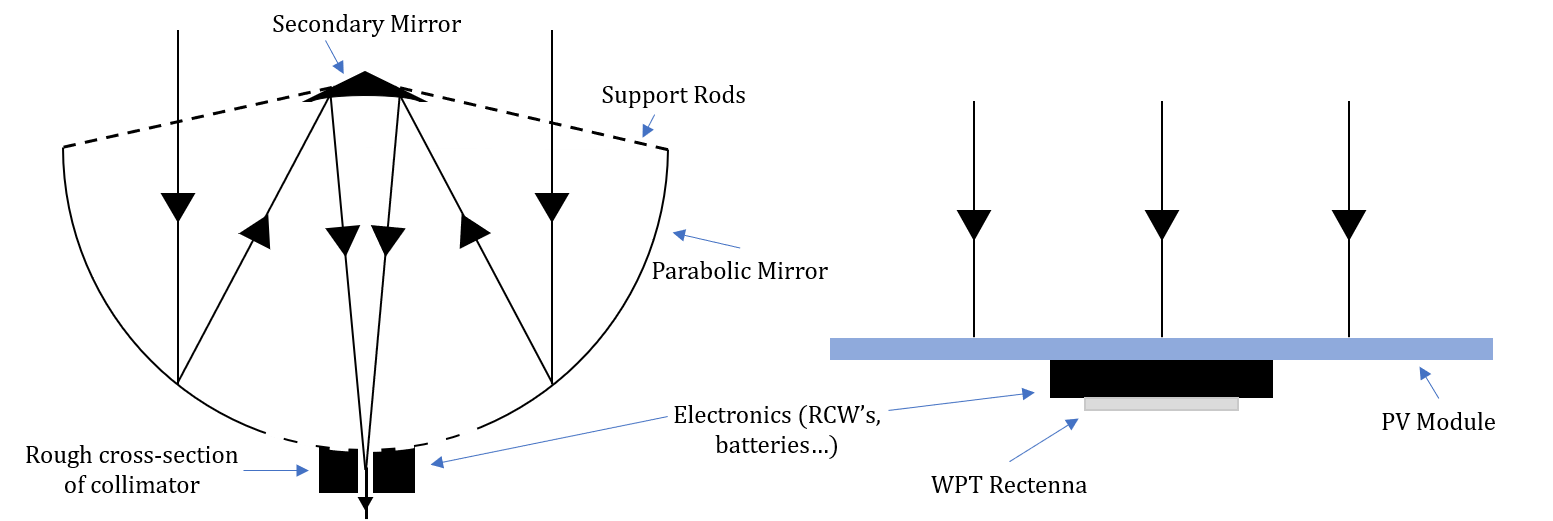}
                \caption{Diagrams of the designs for both the parabolic mirror (left) and PV module (right) approaches. The `electronics' compartments would contain the reaction control wheels, batteries, the electronics which run and control the satellite, etc.}
                \label{fig:twodesigns}
            \end{figure}

        A rough diagram of both designs can be seen in Fig.~\ref{fig:twodesigns} and clearly shows the differences in shape of the two approaches. This is significant because dependant on the method of launching the swarm from Mars, it may be advantageous to have smaller, more compact satellites of which large numbers can be launched at once, or in quick succession. However, the parabolic mirrors could be made more compact for launch by use of an origami-style folding design which can unfold in space, such as those developed by NASA\cite{NASA2017}. At the same time, due to the greatly reduced atmosphere around Mars, air resistance is not a significant obstruction to launching irregular shaped objects. Overall, due to the simplicity of the design of the parabolic mirrors and their ease of production compared to the numerous processes and amounts of materials \& infrastructure required for the manufacture of solar cells, parabolic mirrors in conjunction with concentrated solar power approach is much more desirable due to its higher efficiency, ease of production, and extremely simple satellite design. 
    
\section{Implementation and Construction}
\label{section:3}

        With the design of the swarm itself consolidated, focus can now turn to the process of planning, producing and implementing the Dyson Swarm. As discussed previously, most of the work on the Martian surface will be automated by robots which will be powered by the swarm itself. The types of robots needed are:
        
            \begin{itemize}
                \item Miners | to mine the crust of the planet and obtain the materials needed for construction and fabrication
                \item Transporters | vehicles which transport materials or fabricated parts between locations
                \item Refiners | machines which convert the raw materials obtained by the miners into elements and materials for use in construction
                \item Fabricators | various machines which can fabricate the required parts of the swarm (parabolic mirrors, dielectric coatings, electronics etc.) as well as structures such as shelters from Martian weather (for both humans and robots), launch equipment, storage facilities and more robots.
            \end{itemize}

        The first set of rockets which would be sent to Mars, remembering the two- or so-year gap between launch windows, would likely comprise a few of each of the robots noted above to start a small-scale production process. Initially, these robots will have their own power sources such as solar cells or radioisotope thermoelectric generators (RTG’s), and the fabricating robots will primarily be focused on constructing structures to set up a `base’. Over the two-year period during which these robots are alone, living spaces for humans will be made as well as storage facilities, which will start to be filled from the materials refined that are not yet being utilised. 
        
        With the second launch window, a few humans could be sent alongside more robots and materials to oversee progress and provide means of completing complex tasks and repairing robots/machines. The extra materials sent may be materials which are scarce or not present in the Martian crust and will be needed for the production of the swarm. The robots sent in this launch window will again have their own power sources, but the fabricators will now be capable of producing the various components, materials and objects needed for the swarm. Other structures such as the means of launching the satellites, energy storage and rectenna arrays will be produced over the following two years whilst the production of the satellites themselves will begin, aided by the human workforce. Of course, human presence could be removed, however, it would be beneficial since we are more dexterous and responsive to external factors. The first objects in the swarm will be slowly launched during this period, with the energy produced being kept on Mars to prepare to power the robots produced on Mars, which do not have their own power sources, though this so-called `diversion' of energy from the swarm back to development on Mars will be discussed in greater detail later in section~\ref{section:scale}. The third launch window is similar to the second; it will consist of scarce materials, more self-powered robots and crew exchanges, if required. At this point, the new robots will be configured to produce more robots such that the expansion of the swarm and the robot population can increase self-sufficiently. Over this third two-year period, the swarm will continue to be built and launched whilst more robots are fabricated and powered by the swarm itself, either through charging stations, or possibly even ones utilising WPT technology to enable continual charging of vehicles and machinery. At the same time, infrastructure on Earth will be prepared to receive the energy produced by the swarm from Mars. The biannual launch windows will likely continue to be utilised in order to send crews, materials and other required parts or objects to Mars while the size of the swarm and its production machinery grows independently. With this basic approach laid out, we can now go into more detail about the various sections of production.

    \subsection{Construction}
    \label{section:construc}
    
        When launching the satellites, restrictions on their size will be present; a rocket, for example, typically has a payload enclosed within a cylindrical-shaped fairing. Fuelled rockets, however, are not an ideal method of launching a significantly large number of objects, especially on Mars, so an electric launch system is assumed for the following discussion, the details of which follow in great detail in section~\ref{section:launch}, with the main principle being a tube which accelerates payloads through EM fields, the radius of which can be preliminarily set to 30 cm (with no significant length restrictions), though would naturally vary with the technology and specification of the launch equipment. As such, all components of the satellites, optimally compacted, would need to fit into such an area. All electronics on onboard will need to be kept at operating temperatures through use of the heat produced from the reflected radiation and cooling radiators. For this reason, the electronics forming the satellite bus would be in a casing | with the collimator at the centre | with other components of the satellites being attached on the outside, such as any radiator panels or communications equipment. The collimator itself would be reasonably large; previously a 15 cm radius collimator achieved a divergence half-angle of 1.2°\cite{Fontani2013}. Whilst technology has no doubt improved since then, this 1.2° divergence angle can be used as a reference and its implications will be discussed later. The mass of the satellite bus can be approximated through comparison to those launched previously, such as South Korea’s 200 kg SI-200 bus\cite{SI200}. This bus has many similar components to our design such as thermal control systems, flight software and attitude control, as well as some which we do not require. However, since the collimator and secondary mirror supports must also be considered for our design as well as any other possible design considerations hitherto unconsidered, this 200 kg is a reasonable estimate.

        For the orbital parabolic mirrors, it is first useful to consolidate the materials from which they will be constructed, before finding a constraint for the overall size and shape of the payload which will be placed into orbit. As outlined in section~\ref{section:reflectingsheets}, the main structure of the dish will be metallic with a dielectric coating, the latter of which doesn’t have a significant mass and can be accounted for by using slight over-estimates for the mass of the metallic structure beneath it. The mass of both the satellites and the materials required for construction can be reduced by using a metal with a lower density, which of the iron, aluminium and magnesium identified as prime candidates earlier, have densities of 7.75, 2.70 and 1.78 g/cm$^3$. These metals alone or alloyed together (possibly including with other elements available in Mars’ crust) would allow for a lightweight support for the secondary mirror, not to mention for any other supports and casings on the satellites. The rod-like support for the secondary mirror could be from the centre of the bus of the satellite, however, since this is also where the collimator is located, one option is to reflect the radiation from the secondary mirror into its support rod | which could be, or could lead to the collimator | however, this would signify the swarm being in areostationary orbit (20.4 km), which is rather close to the moon Deimos (23.4 km | a half tonne mass at this separation would experience a force of 5.47 N). An alternative is to support the secondary mirror from multiple (likely three) separate points at the edge of the primary mirror which, whilst blocking a small proportion of light from reaching the primary mirror, would result in a higher overall efficiency and the effect of two extra supports on the net mass of the system is negligible. This would likely require a mechanical or at least mobile component as the supports would have to move and connect to the secondary mirror once in orbit, hence supporting the secondary mirror from the edges of the mirrors is ideal. 
        
            \begin{wrapfigure}{r}{0.4\textwidth}
                \centering
                \includegraphics[width=0.25\textwidth]{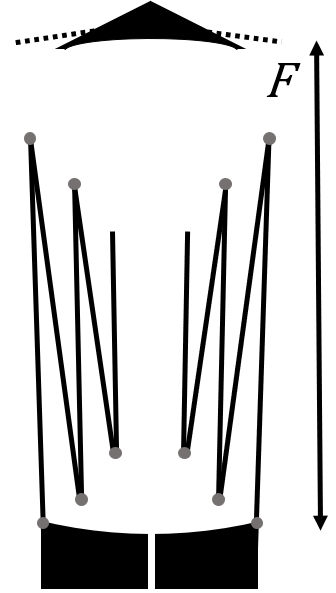}
                \caption{Crude illustration of a folded parabolic mirror satellite compacted for launch with focal length, $F$. Utilisation of the Miura fold would compact the individual sheets even further, but this is not illustrated for clarity. Similarly, the rods which support the secondary mirror from the edge of the unfolded primary mirror are shown as dotted lines in the locations they would be for simplicity.}
                \label{fig:folddesign}
            \end{wrapfigure}
            
\begin{figure}[h]
                \centering
                \includegraphics[width=0.6\textwidth]{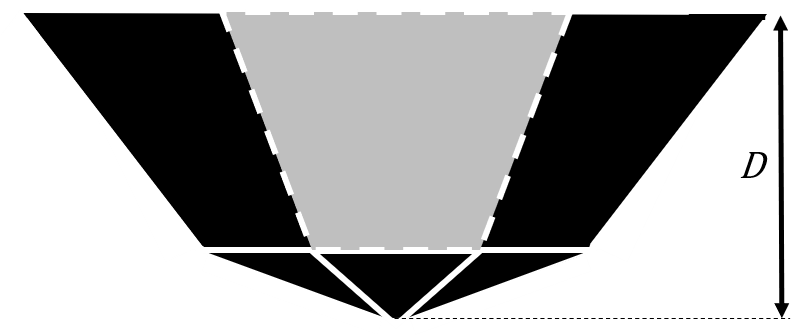}
                \caption{Diagram of a mesh for a parabolic mirror formed of planar sheets with depth, $D$. Diagram modelled off Fig. 1 of \cite{Jape2020}.}
                \label{fig:parabolaapprox}
            \end{figure}

        For a rough estimate of a density and mass, the densities of each of the aforementioned metals can be combined by their abundances in the crust, which results in a benchmark density of 5.335 g/cm$^3$. The mass of the mirrors can be estimated using their thickness (assumed here to have an average value of 5 cm) and their radii, which can now be addressed. Since the radius of the primary mirror will likely be larger than the 30 cm limit, they can use principles of origami to fold up into a smaller area for launch, before unfurling when in orbit, like NASA’s James Webb Space Telescope. One such concept is the Miura fold, which allows a flat sheet to be folded into a much more compact shape, whilst being able to unfold in one movement\cite{Nishiyama2012}. Clearly parabolic mirrors are not flat plates; however, it can be noted that the radiation does not need to be concentrated upon a particular point after reflection from the secondary mirror | just into the collimator, the opening of which can be made funnel-like. This implies that both the primary and secondary mirrors do not need to be perfect parabolas but can approximate such a shape via a collection of flat plates at angles to one another, still resulting in the light being reflected into the collimator. This design, as shown in Fig.~\ref{fig:parabolaapprox} would allow the primary mirror to be folded up in a much easier fashion than previously illustrated, and vitally take up less space. Fig.~\ref{fig:folddesign} shows the cross-section of a folded satellite with focal length, $F$, in a simplified manner. Assuming that any sized primary mirror would be able to be folded into a cylindrical shape of radius 30 cm in the length of its own focal length, the size and mass of both mirrors can now be determined. 
        
        Due to the approximated parabolic shape of the primary mirror, the diameter of the secondary one would have to be at least the same length as the longest of the panels which make up the parabolic primary mirror, since then all of the reflected light from any primary mirror panel will be incident upon it the secondary mirror. This particular value has no significant effects on the satellites, so a 0.25 m radius secondary mirror is chosen, which is small enough to fit within the size limitation, but these mirrors could equally be folded if needed. It should be noted that regardless of the radii of both mirrors, to achieve the same total amount of energy from a swarm, the same amount of material is still required to build all of the satellites. Larger mirrors do, however, mean that fewer objects would need to be built in total as each would have a larger power output, but this also introduces issues of transporting and launching such large structures, thus a compromise can be found in between. As such, the radius of the primary mirrors can also be arbitrarily chosen and a 10 m radius is selected such that the satellites will be relatively small when compacted yet have a significant aperture of 314 m, corresponding to 122 kW of solar radiation. At any given radius, however, the focal length, $F$, and depth, $D$, of the primary mirror do affect the mass of the mirrors and are related to the radius, $R$, by:
        
            \begin{equation*}
                4FD=R^2
            \end{equation*}

        More specifically, the increase in mass required is inversely proportional to the FD ratio ($F/D$); a larger FD ratio is henceforth ideal. Due to the nature of the dual-mirror design, the focal lengths of both mirrors are the same, with the assumption that the secondary mirror’s focal point is the centre of the primary mirror’s base. With this | and the radius of both mirrors | in mind, a focal length and depths for both can now be decided upon. A larger focal length reduces the total mass of the mirror, yet one too long is physically unreasonable, since this increases the lengths of the secondary mirror supports. A benchmark of a ten-metre focal length is considered with the implied changes in depth and mass of both mirrors shown in Table~\ref{table:focal}. Six metres is the lower bound for the focal length, since below this value, the focal length becomes smaller than the depth.

            \begin{table}[h]
                \centering
                \caption{Pairs of values of the focal length, $F$, and depth, $D_p$, of the primary mirror from a fixed radius of 10 m. Each is accompanied by the implied depth, $D_s$ of the secondary mirror and the combined mass of both mirrors, $M_p+M_s$.}
                \label{table:focal}
                \begin{tabular}{cccc}
                \hline \hline
                \textbf{$F$ (m)} & \textbf{$D_p$ (m)} & \textbf{$D_s$ (mm)} & \textbf{$M_p+M_s$ (kg)} \\ \hline
                13 & 1.9 & 1.2 & 87.5 \\
                12 & 2.1 & 1.3 & 88.0 \\
                11 & 2.3 & 1.4 & 88.6 \\
                10 & 2.5 & 1.6 & 89.4 \\
                9 & 2.8 & 1.7 & 90.5 \\
                8 & 3.1 & 2.0 & 92.0 \\
                7 & 3.6 & 2.2 & 94.2 \\
                6 & 4.2 & 2.6 & 97.5
                \end{tabular}
            \end{table}
            
        Clearly an 11\% increase in mass when moving from $F=13$ m to 6 m is ideally avoidable in the long run so a compromise between length and mass can be struck at a focal length of around 10 m, which has no significant increase in mass from $F=13$ m but helps reduce the lengths and thus stresses on the secondary mirror supports. The implied depth of the secondary mirror seems small, at 1.56 mm, however, such a small depth can be achieved more easily with the flat panels outlined above, especially since the target from the secondary mirror is not point-like. The depth of the secondary mirror can be increased by simply increasing its radius at constant focal length. With these parameters for both of the mirrors, their final specifications can be laid out. Primary mirror: $R=10$ m, $D=2.5$ m, $F=10$ m, $M=89.4$ kg. Secondary: $R=0.25$ m, $D=1.42$ mm, $F=10$ m, $M=52$ g. The implications of these parameters on the overall power output of the swarm will be discussed in section~\ref{section:4}. The mass of a single satellite within the swarm can be estimated by adding the $\sim$90 kg mass of the mirrors to that of the satellite bus to gain a rough value of 290 kg. Clearly it must be ensured that the satellites would be able to withstand the forces and disturbances (vibrational, magnetic etc.) during launch through prior design and testing on Earth. 
        
    \subsection{Swarm Orbit}
    \label{section:orbit}
    
\begin{figure}[ht]
              \centering
              \begin{subfigure}[b]{0.51\textwidth}
                 \centering
                 \includegraphics[width=\textwidth]{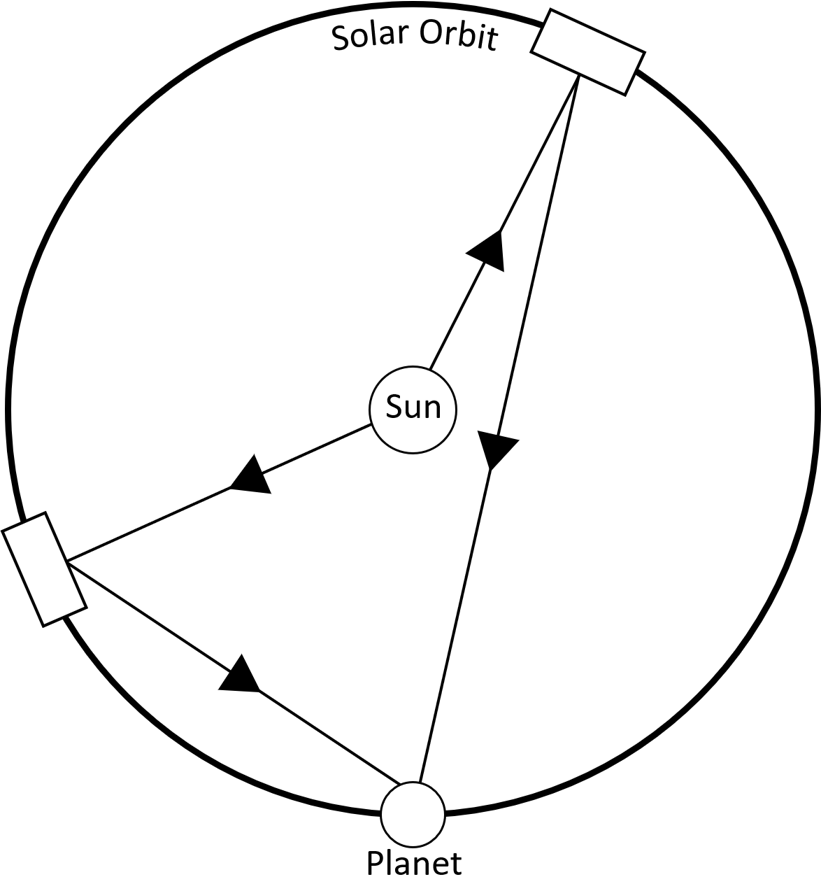}
                 \caption{Solar Orbit}
                 \label{fig:solarorbit}
             \end{subfigure}
             \hfill
             \begin{subfigure}[b]{0.48\textwidth}
                 \centering
                 \includegraphics[width=\textwidth]{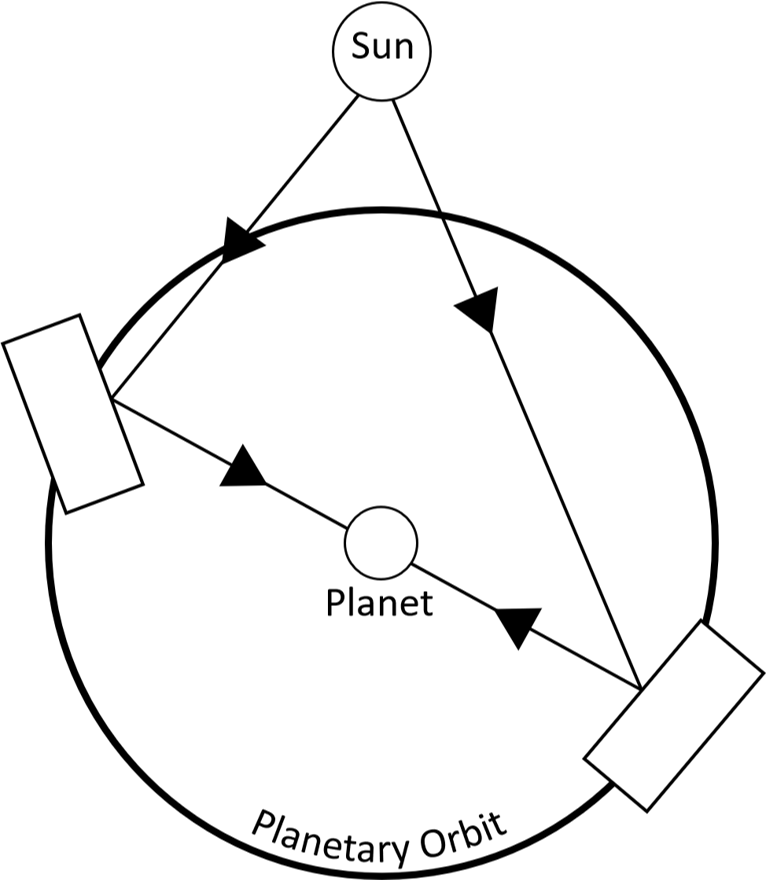}
                 \caption{Planetary Orbit}
                 \label{fig:planetaryorbit}
             \end{subfigure}
                \caption{Diagrams showing the differences in the paths of radiation taken from the Sun to the host planet via the rectangular Dyson Swarm satellites which are deployed in either a solar orbit (a), or a planetary orbit (b).}
                \label{fig:orbitcomparison}
        \end{figure}

        As implied throughout this discussion, for the swarm’s specific location with regard to Mars, it is beneficial to place the swarm in orbit about the planet versus in solar orbit. Firstly, this allows us to decrease the amount of debris in space which is ideal to prevent collisions with other satellites and spacecraft, as well as reducing the impact of the swarm on the night sky for astronomy and our species. Secondly, by keeping the distance between the host planet and the satellites at a minimum, the radiation has to travel a shorter distance, meaning that the efficiency of the transmission of energy between the objects and surface collectors is kept as high as possible. As illustrated in Fig.~\ref{fig:solarorbit}, radiation must travel the distance of the orbital radius of the swarm and then an extra distance to reach the host planet, making the energy obtained by those the farthest away almost negligible. The objects in Fig.~\ref{fig:planetaryorbit}, however, have to traverse the same orbital radius but then the average distance between the objects and the planet is simply the height of the orbit of the swarm. The latter is more efficient with the same number of objects. This does mean that there would need to be numerous stations as the radiation cannot penetrate the planet itself, but a similar problem occurs with a swarm in solar orbit, since the swarm would only be able to transmit to the side of the planet visible to them, which would not always be the same, thus in a similar fashion, numerous stations would need to be built around the planet, though not all would be used simultaneously.
        
        The height to which the swarm will be launched should be considered to understand the launch requirements for this orbit. Orbital radii near those of both of Mars’ moons (9.3 km and 23.4 km) will be avoided to prevent collisions and gravitational interference. Similarly, despite Mars’ very thin atmosphere, dust particles are common in the troposphere (below 40 km), which could land on and reduce the reflectivity of the parabolic mirrors and could also degrade them\cite{HABERLE2015}. The top of the troposphere is also home to water and \ch{CO2} ice clouds, which should be avoided. A higher orbit will also signify a lower orbital velocity, which could make launching the swarm easier, however, an increased distance increases dispersion of the reflected radiation from the swarm. Thus, it is optimal to place the swarm above the troposphere and a 55 km orbit is chosen which would produce an orbital velocity of 3520 m/s and an orbital period of 1 hours 43 minutes, whilst remaining above the Martian dust and clouds. Such a reasonably short orbital period will mean that the satellites are only exposed to solar radiation for around 52 minutes before they transit the far side of the planet, which additionally may help with any cooling needs.
        
        The circular mirror array on the surface of Mars would be formed by numerous mirrors of the same specification as those being launched into orbit, but of the required shape to reflect the radiation onto the solar cells or generators, for example the flat, rectangular heliostats in Fig.~\ref{fig:CSPTypes} or possibly smaller parabolic mirrors. The size of the solar array or generator would simply depend on how capable they are at converting all of the incident radiation to electricity. The size of the mirrors can be chosen arbitrarily to cover the area over which the reflected radiation from the swarm will be incident, a smaller mirror size would only signify that more mirrors would make up the array, and vice versa. The area over which the incident radiation is spread can be found from the estimated divergence of the collimator. The 1.2° degrees divergence half angle and the orbital height of 55 km would mean that the radius of the mirror array would need to be about 1.15 km | an area of 4.17 km$^2$, which has already been achieved on Earth by one of the three heliostat arrays at the Ivanpah Solar Power Facility, USA, which has a radius of 1.75 km\cite{Sullivan2015}. This 1.15 km may increase slightly to account for the multiple orbital planes being marginally off from the zenith of the collecting stations, but this increased radius is unlikely to exceed 1.5 km. The mirrors would not be able to completely reflect all of the light coming from the swarm due to the spacing between them, which would depend on the dimensions of the solar array/receiving engine. The power collected from the collecting stations on Mars would be sent directly to either the WPT transmitters sending power to Earth, or elsewhere for use on Mars itself.
        
    \subsection{Launch}
    \label{section:launch}
    
        The method of launching the satellites into orbit must be easy, energy conservative and capable of launching with small intervals. Typical rockets are not ideal for use since they rely on fuel and would contribute large amounts of space debris unless they were recovered, though an exception to this is deploying some of the satellites into orbit when rockets are sent back to Earth with crew, but this comes with its own problems, the most obvious of which is that this event only occurs every two years at most. Therefore, an optimal launch method would use the electricity produced by the swarm itself. Electric or ion engines are not appropriate for this purpose since they require a gas propellant, aren’t re-usable, and more vitally, would struggle to lift payloads off the ground, as they provide low thrusts over long times to make up high $\Delta V$’s. Alternatively, a type of electromagnetic railgun may be used, in which a magnetic field could accelerate individual projectiles to high speeds from the ground and launch them into orbit. A mass driver is of a similar design to an EM railgun but instead uses superconducting magnetic coils to accelerate a payload down a long track before release\cite{MassDriv}. Non-superconducting (resistive coils) can also be used, but this significantly increases the operating energies required\cite{ROSHAL2018}.
        
        A benchmark to the power and scale of a mass driver was proposed by Henry Kolm, one of the leading workers on the idea at the time\cite{MassDriv}. He describes a telephone pole-shaped vehicle which would sit inside a mass driver with a 12’’ (30 cm) radius and 7.8 km length. The length of the system here reflects a vertical mass launcher to the deepest well drilled on Earth, projectiles of which would experience minimal atmospheric disturbances, though on Mars of course, this is not a significant issue. However, in the context of this Earth-based mass driver, the one tonne telephone pole would have a launch velocity of 12.3 km/s, having accelerated at one thousand gees for 1.26 seconds and consumed an average power of 60 GW. These impressive numbers are in excess of what would be needed on Mars | escape velocity for the planet is only 5.03 km/s and as we only wish to get into orbit, we can put the required $\Delta V$ at 3.52 km/s (for a 55 km orbital radius). By recalculating these values with this new final velocity parameter at the same acceleration, we obtain:
        
            \begin{equation*}
                v=u+at \Rightarrow t = \frac{v-u}{a} = \frac{\Delta V }{1000\cdot g} = 0.36 \text{ s}
            \end{equation*}
            \begin{equation*}
                v^2=u^2+2as \Rightarrow s = \frac{v^2-u^2}{2a} = \frac{\left(\Delta V\right)^2}{2\cdot1000\cdot g} = 631 \text{ m}
            \end{equation*}

        which is a consequential decrease from an Earth-based system. The larger power consumption is an issue which can be addressed by increasing the length of the launcher. It should also be noted that the launcher would not need to be built vertically into the crust of Mars, since the lack of an atmosphere removes the reduced efficiencies of launching at an angle, so the launcher would be built on the surface, possibly along a hill or mountain to provide the slope to reach the desired orbit. Since the power and distance are linearly related, doubling the length of the launcher will halve the power required; the same effect as halving the mass. Using the estimated mass of $\sim$290 kg from the previous section and increasing the length to 10 km would signify a power of 871 MW, which is significant to begin with, but would be a negligible amount once the swarm reaches operational sizes. To power the mass driver in the infancy stages of the swarm’s production, either other methods of power generation or a different process of launching the satellites would be required. If the entire swarm is launched using just one mass driver, they will all be in a similar orbital plane, as mentioned earlier. Depending on the numerical size of the swarm, however, different orbital rings will need to be utilised each of which would likely require a new ring of receiving stations and a new mass driver (unless the previous one is moved/reorientated). Discussion of the WPT system below covers the orbits into which the swarm will be placed, with the final layout determined in section~\ref{section:scale}.
 
    \subsection{Transfer}
    \label{section:transfer}
    
\begin{figure}[h]
                \centering
                \includegraphics[width=0.75\textwidth]{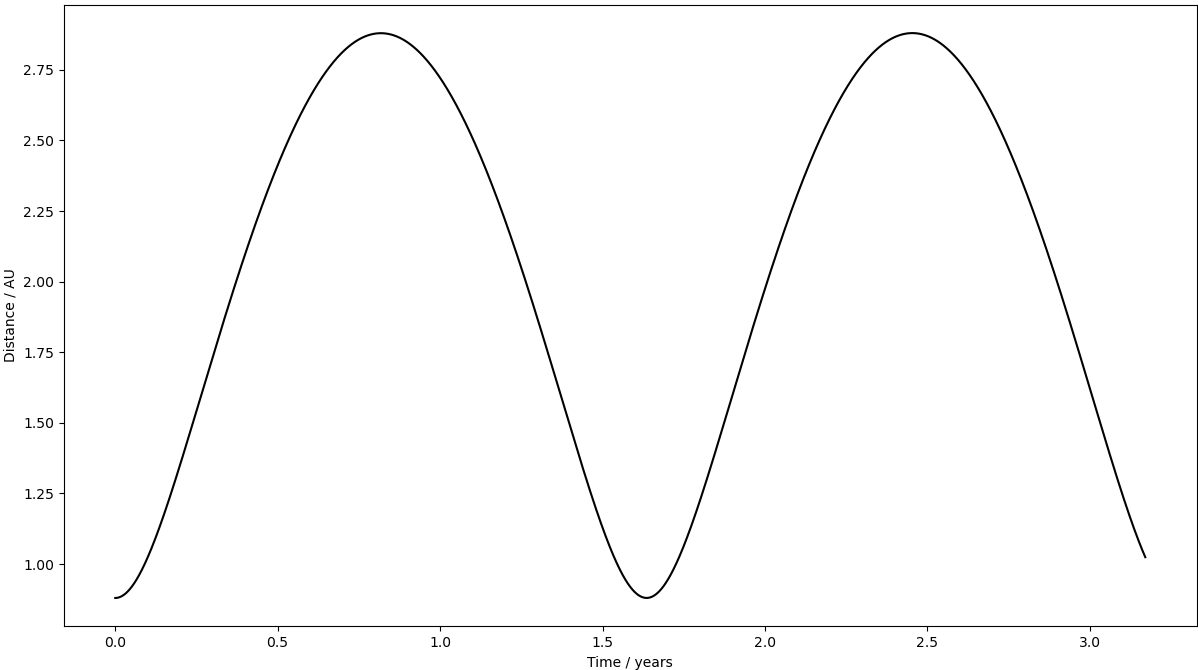}
                \caption{Graph of the distance between Earth and Mars over the course of three years, from a simulation of the inner planets. The optimal transmission time occurs about every 1.7 years, with a reasonably sized window lasting for 2 - 3 months either side.}
                \label{fig:marsearthdist}
            \end{figure}

        As discussed, wireless power transfer will account for the transmission of the power obtained from Mars to Earth. To send energy back to Earth, it is impractical to transmit directly from the surface of Mars to Earth, since microwaves are scattered by water droplets in the Earth’s atmosphere. Instead, transmission from Mars could be received by one or more orbiting stations around the Earth. Due to the varying distances between Mars and Earth, the efficiency of the transmissions will deteriorate as this distance increases. Depending on the power output of the swarm, this may be acceptable, however, in its initial stages the power obtained from the swarm could be stored on the surface of Mars until the distance between Mars and Earth is a minimum, which would be a discrete period biannually. The period can be a large as desired, though the efficiency of the transmission would only peak at closest approach, as shown through Fig.~\ref{fig:marsearthdist}. The Earth-orbiting relays are unlikely to have their own capacities, since the amounts of energy being transmitted would require immensely large storage facilities. The location of the ground receivers on Earth would be determined by wanting to have the orbiting transmitters at the zenith to maintain a high efficiency transmission, whilst also having locations with low precipitable water vapour (PWV) values. For example, ideal ground sites would be at latitudes above 40°N and below 40°S, where the mean global PWV is less than 20 mm, as opposed to up to 50 mm at the equator\cite{Chen2016}. In turn, this determines the orbital planes into which the orbiting receivers/transmitters must be launched. 
        
        The receiver(s) in orbit around Earth which receive transmissions from the Martian surface could be placed in a Sun-synchronous orbit (SSO) to ensure that they will always be above the ground station on Earth to which it will transmit, allowing for transmissions to be made when weather and PWV conditions are optimal. The use of multiple SSO satellite receivers would ensure that one is always in contact with the Earth ground station, and so continuous transmission from Mars to Earth would be possible. An SSO is better than a geostationary orbit, since the latitude of the ground station is not restricted to the equatorial plane at which PWV’s are higher. Transmissions between the two planets will likely only occur when the Earth-based SSO satellite is not in conjunction with the Earth itself. This ensures that any transmitted radiation which may miss the receiver is not incident onto the Earth’s surface. As there will be numerous stations on Mars for different parts of the swarm to reflect radiation to, there will similarly be more than one WPT transmitter on the ground for continuous transmission to Earth. Both transmitters (on Mars \& in orbit) may utilise phased arrays (antenna arrays which allows for directional transmission) to ensure their transmissions are always in the direction of their target or, for the case of Mars to Earth, where their target will be. 
        
        The efficiency of the WPT systems is dependant on the four main components of the process, as shown below, with their typical efficiency ranges shown in parentheses; the exact specification of the components will determine their efficiencies, so typical values are used as a guide\cite{Brown1992}:

            \begin{itemize}
                \item Converting stored energy to microwaves (70 - 90\%)
                \item Forming a beam via an antenna (70 - 97\%)
                \item Transmission of the beam (|)
                \item Conversion of the received beam back to energy (85 - 92\%)
            \end{itemize}
                    
        Each of the given ranges are quite acceptable for this application and may improve further in the coming years. The efficiency of the transmission of the microwave beam itself, however, is dependant on a range of factors which will be unique to the parameters of each transmitter and receiver. To obtain an estimate for the efficiency of such transmissions, these parameters of the design of the WPT system must be decided upon, namely, the size of the transmitting and receiving antennae, the distance between them, $D$, and the wavelengths of the microwaves being transmitted, $\lambda$. These parameters are linked to the parameter $\tau$ by the following equation\cite{Brown1992}:
        
            \begin{equation*}
                \tau^2=\frac{A_tA_r}{\lambda^2D^2}
            \end{equation*}

        where $A_t\text{ and }A_r$ are the aperture areas of the transmitting and receiving antennae, respectively. The easiest of the variables to decide upon is the wavelength of the microwaves. This is because a smaller wavelength will result in a higher value of $\tau$ and thus a higher efficiency, $\eta$, due to the relation:
        
            \begin{equation*}
                \eta=-e^{-\tau^2}
            \end{equation*}

        The microwave range has a typical lower bound of around one millimetre, which is largely ideal for Mars as there are no significant absorbing mediums. Whilst antenna aperture areas (also called \textit{effective area}, $A_e$) are not directly related to the physical aperture size, $A_{phys}$, the aperture efficiency, $e_a$, relates both quantities:
        
            \begin{equation*}
                e_a=\frac{A_e}{A_{phys}}
            \end{equation*}

        Typical antennae have aperture efficiencies of around 70\%, which allows us to estimate an effective area for a given physical aperture size\cite{Vandelle2018}. Both transmitting and receiving antennae on Mars, Earth and orbit can be as large as desired within mechanical and constructional limitations. They can take the form of antenna arrays to cover a large area, such as the initial transmission from Mars, or parabolic antenna dishes, which may be used on Earth to receive the final transmission from orbit. The specific design doesn’t affect the transmissions, as the efficiencies are only dependant on the apertures of the equipment. Firstly considering the transmission from Mars to Earth orbit, a preliminary choice for the sizes of the transmitting and receiving arrays is a $1000\times1000$ m (On Mars) and $100\times100$ m (Earth orbit); effective areas of 1 km$^2$ and 0.01 km$^2$, respectively. To complete the analysis of the microwave transmission efficiency, the distance between the two antennae are considered; the closer the paired antennae are, the higher the efficiency. Two values will be used to show the differences between transmitted only when Mars and Earth are closest (0.36 AU), and continuously, regardless of their separation (average of 1.50 AU). With even the smaller of these two distances, the parameter $\tau$ is expressed as:

            \begin{equation*}
                \tau^2=\frac{0.7^2\cdot\left(1\times10^6\cdot0.01\times10^6\right) }{0.001^2\cdot\left(0.36\text{ AU}\right)^2} = 0.00000169
            \end{equation*}

        which leads to an extremely low efficiency of 0.00017\%. Clearly much larger apertures are needed; of course an optimal efficiency would be 99.9\%, however, for a 5000 metres squared receiver the Martian transmitter would have to be of an area around 1600 km$^2$, just bigger than the land area of the Faroe Islands. This is a significant size | though not completely unexpected due to the scale of the project | which will likely decrease over time as technology advances, such as aperture efficiencies reaching closer to 100\%. The receiver could be made possible by using numerous antenna arrays, which may be folded for launch and joined together when expanded in orbit to form larger apertures. Higher frequency waves could also be transmitted such as those in the infrared: a wavelength of 10 µm would only need transmitting and receiving apertures of around 28 km$^2$ and 0.13 km$^2$, respectively, for an efficiency of 99.8\%. This is a significant decrease from higher wavelengths and could be promising, and a slight overestimate of aperture sizes of 30 km$^2$ and 0.15 km$^2$ are used as an estimation for this 99.8\% efficiency. It is likely that at least two, and possibly three of these transmitters would need to be constructed on Mars, such that one is always in contact with the SSO satellites around Earth, regardless of Mars’ orientation. The exact location of the transmitters on the surface is not too specific, though the topology of the ground may help with their deployment and function; the only restriction is to ensure that the path of the transmitted microwaves is clear and does not intersect with the orbital swarm. This is accomplished by choosing the location of the WPT transmitters to be on a different plane to any of the collecting stations and swarm bands | these `bands' would be orbital rings of the swarm satellites, which would ensure an orderly deployment of the swarm and reduce debris.
        
\begin{figure}[h]
                \centering
                \includegraphics[width=0.7\textwidth]{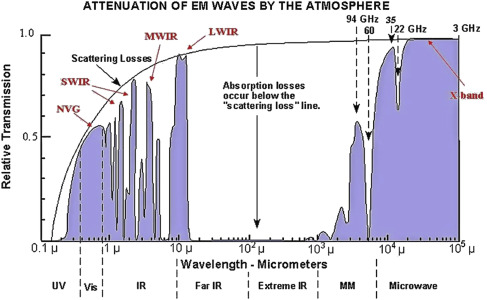}
                \caption{Graph of the relative transmission of wavelengths of the EM spectrum through the Earth's atmosphere\cite{FINGAS2017}.}
                \label{fig:windows}
            \end{figure}

        With such an efficiency for the transmission of the microwave beam, the other three stages of the WPT process can be combined to gain an idea of the overall efficiency of the transfer. Using the range of efficiencies given by Brown, the efficiency of the WPT system as a whole ($\lambda=10$ µm, $D=0.36$ AU, $A_t=30$ km$^2$, $A_r=0.15$ km$^2$) is given by the product of the constituent efficiencies and would range from 41.6 – 80.1\%. A similar analysis can be applied to the other transmission stage; transmitting from the satellites in SSO to the ground station, the reduced transmission distance reduces the required sizes of the antennae. The wavelength of the microwaves in this case has an ideal value of 1.26 cm, since this corresponds to the X band transmission window, as shown in Fig.~\ref{fig:windows}, allowing for an efficient transfer in terms of atmospheric absorption and scattering\cite{FINGAS2017}. The increased wavelength results in a pairing of a roughly 2000 m$^2$ transmitting antenna and a 0.4 km$^2$ receiving array on Earth, which would provide a 99.8\% efficient transmission (excluding atmospheric attenuation). The specific design of these SSO satellites is not discussed mainly because they will be constructed and launched from Earth, which is significantly easier than from Mars, thus their designs are less constricted and could follow numerous approaches.  
        
        By optimising the size of the effective areas of the transmitting and receiving antennae, both transmissions can be optimised, with the total efficiency of the WPT transfer from start to finish (based purely on the efficiency of the WPT components) ranging from 17.3 – 64.3\%, depending on the end efficiencies of the converters and beam-forming antennae, though one can assume that the technology used for the swarm would be state of the art at the time of construction to maximise its output, justifying the credibility of higher efficiencies. Considering the intensity of a point source would reduce by about $10^{22}$\% when travelling between Earth and Mars at their closest approach, even a 17\% efficiency is a significant achievement. As outlined earlier, the average separation of the two planets (1.5 AU) must naturally be discussed. With the same aperture sizes and wavelengths of transmission, the efficiency from Mars to Earth orbit is still a surprising 12.2 – 23.5\%, which becomes 5.1 – 18.8\% when the transmission from Earth orbit to the ground is added. This is a good baseline considering the large distances being traversed by the radiation, and one which can only increase with the technological advancements which may come in the future.
    
\section{Energy and Implications}
\label{section:4}

    \subsection{System Efficiency}
    \label{section:systeff}

\begin{figure}[p]
            \vspace*{-2cm}
            \makebox[\linewidth]{
                \includegraphics[width=1.3\linewidth]{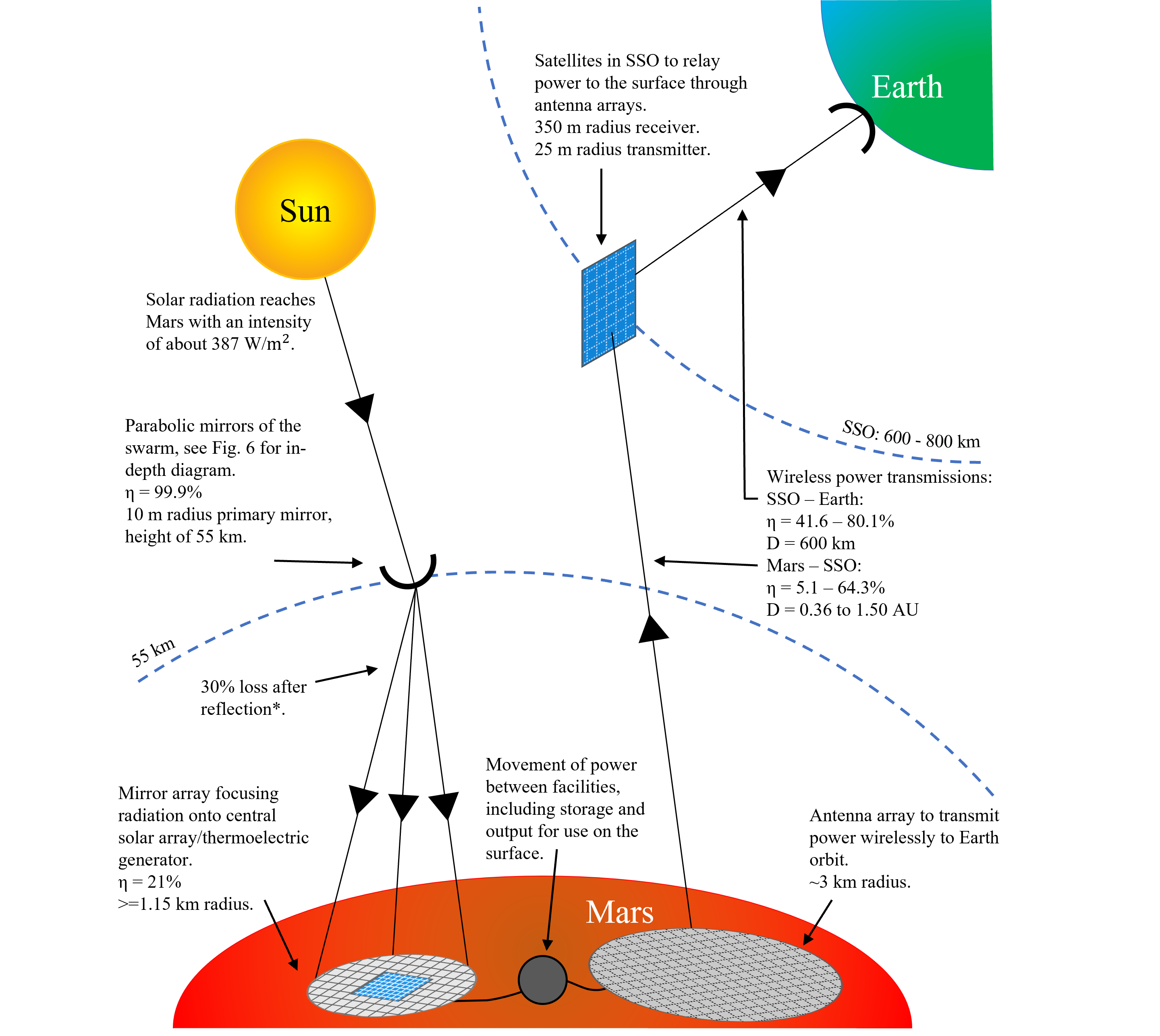}
            }
            \caption{Diagram showing an overview of the journey of radiation from the Sun to Earth through the Mars-based swarm. *Efficiency of transfer from swarm to mirrors can be estimated to be 70\%, including absorption of solar radiation in Mars’ atmosphere, dispersion of reflected radiation, and the land area uncovered by mirrors.}
            \label{fig:system}
        \end{figure}

        With each part of the Dyson Swarm now discussed, an overview of the design can be reviewed. The swarm itself would consist of individual reflecting mirrors which will reflect incident solar radiation to certain points on the surface of Mars at which collecting stations would be located. These will turn the incident radiation to electricity by use of solar cells or thermoelectric generators, as well as an array of mirrors to help maximise the amount of radiation they receive. The generated electricity, other than being used on Mars, will be sent back to Earth by a collection of wireless power transmission stations which convert the electricity to microwaves for transmission to satellites in sun-synchronous orbits which then relay the electricity down to Earth. These processes are illustrated in Fig.~\ref{fig:system} with the efficiencies of each process noted, as well as the sizes of some of the systems' key components. This allows us to gain a first glimpse of the overall efficiency and output of the system, which will allow us to refine the swarm if and as needed.
        
        In the early stages of the swarm’s development, power may be stored on Mars before being transmitted back to Earth when the two planets are at their closest separation of 0.36 AU, at which distance the total efficiency of the swarm ranges from 2.54\% to 9.44\%; the product of the efficiencies of each of the separate processes the radiation is subjected to. As the size and capability of the swarm increases over time, power will likely be transmitted to Earth continuously, with the average separation of 1.5 AU over each two-year period at the opposition of both planets. At this average distance (which will be used from now on when referring to the swarm’s efficiency as continuous transmission would occur for a large swarm), the efficiency reduces to 0.74 – 2.77\%. As one would expect due to the vast distances involved, this is a small value, however, it should be noted that even if Mercury were used due to its closer proximity to Earth, the efficiency would still only range from 2.38 – 8.85\%. 
        
    \clearpage    
        
    \subsection{System Scale}
    \label{section:scale}
    
\begin{figure}[h]
                \centering
                \includegraphics[width=\textwidth]{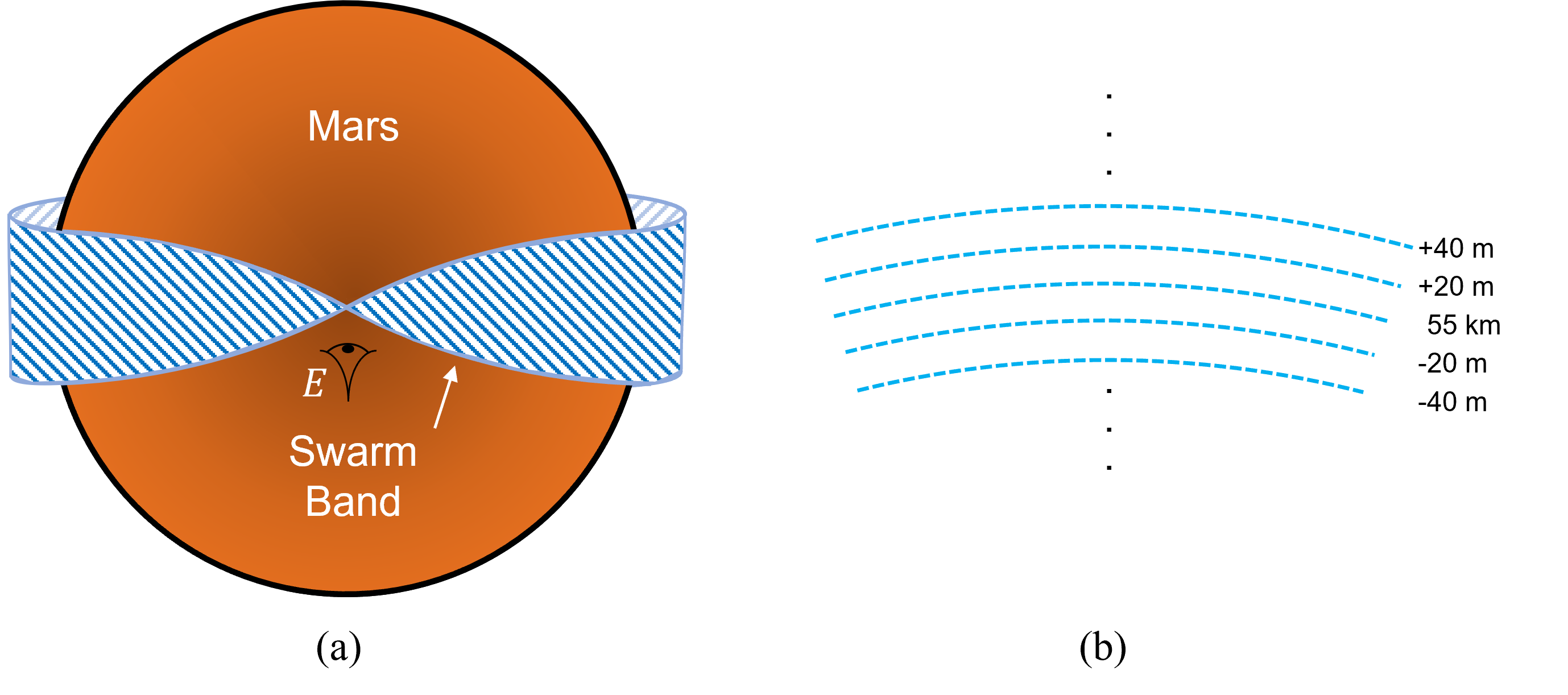}
                \caption{A rough diagram to show the concept of combining multiple orbital planes into a so-called swarm `band'. Figure (a) shows the swarm around Mars, and (b) the side-view from point E in (a) of the constituent orbital planes.}
                \label{fig:orbits}
            \end{figure}

        The scale of the Dyson Swarm can be analysed by using a benchmark power output which we would like the swarm to have. Naturally, the power output of the swarm will be less than its complete capacity, since some of the electricity produced will be used to power some of the swarm’s infrastructure, like the 871 MW mass driver, as well as all of the robots, buildings etc. on Mars itself. An estimate for the amount of power this would consume is difficult to obtain. However, if we take the Earth’s 2019 total energy consumption of $P_C=18.5$ TW from earlier (\cite{BP2020}) the mass driver’s 871 MW consumption alone is a negligible 0.0047\% of $P_C$, suggesting that the overall consumption on Mars will be low. Even if non-negligible, 11.41\% of Earth’s 2019 energy consumption came from renewable or hydroelectric sources thus it can be assumed that the energy consumed by the infrastructure on Mars will either be insignificant or can be counteracted by renewable generation on Earth, so this 18.5 TW is a reasonable value to want to achieve. Throughout these calculations, the upper bound of efficiencies described in section~\ref{section:systeff} to optimistically account for advancements in technology before the time of the swarm’s construction. The number of satellites, N, making up the swarm is given by:
        
            \begin{equation*}
                N=\frac{P_C}{I_{\mars}\eta A_s} = \frac{18.5\times10^{12}}{387\cdot2.77\%\cdot100\pi} = 5.51\times10^9
            \end{equation*}

        where $A_s$ is the aperture size of the primary mirror of the satellite ($\pi r^2$). To fit the satellites into orbital rings, the separation of the bands outlined in section~\ref{section:transfer} must be determined as well as the number of orbital heights within a band. This is illustrated in Fig.~\ref{fig:orbits} (b). The cuboidal volume of space taken up by the satellite itself (radius 10 m) is roughly $20\times20\times12.5$ m, or 5000 m$^3$; an area of $A=20\times20=400$ m$^2$. If all 5.5 billion satellites were spread out at an orbit or 55 km as tightly as possible, they would only cover 1.48\% of the area available around Mars. Adding a minimum spacing of 20 m between each satellite to reduce the chance of collisions and to prevent occultations of satellites would effectively increase the area of one satellite to $A=40\times40=1600$ m$^2$ which in turn increases the percentage of the orbital area required to be covered to 5.91\%; a large yet physically reasonable size. 
        
        Spreading the swarm in orbit around Mars at a single height is not ideal, however, since this will pollute the space around Mars, preventing ships from travelling to and from the surface as well as ruin the view of Mars from Earth. At the same time, a swarm spread out across the whole planet would require countless collecting stations to which the reflected radiation will be directed to. To solve these issues, there are numerous ways of deploying the swarm into orbit. One method is placing the satellites into orbital planes which overlap at a point either side of the planet, meaning only a few collecting stations would need to be built on the ground directly below that orbital plane, and similarly, launching the swarm would be easier since the trajectory to orbit would be the same for each orbital plane. This also reduces the area covered by the swarm since, to avoid collisions at the two point of intersection, the orbital planes would have to be at slightly different heights; this idea is shown again in Figures~\ref{fig:orbits} (a) and (b). There would be a negligibly small loss around the points of intersection where some satellites would occult one another.
    
        Other approaches to achieve the same goal need not be discussed, however, the baseline of the area of a 55 km orbit which would need to be covered is 5.91\% when each of the satellites is given a 20 m buffer zone in both the $x$ and $y$ directions. By placing the swarm over an area this size which is restricted to certain orbital planes, it can be ensured that Mars’ neighbouring space can be kept clear in most places, allowing for other space activities and reducing the visual impact of infrastructure of this scale. Clearly, the consumption of energy on Earth will have increased beyond 18.5 TW by the time the swarm is completed; 2030 consumption is estimated to be $\sim$10\% higher than that of 2019 (20.35 TW)\cite{IEA2020}. The corresponding swarm size is just over 6 billion, which relates to covering 6.49\% of a 55 km high Mars orbit (without using multiple orbital heights). This means that the swarm would be able to cater for the Earth’s increasing energy need with population growth and the energy demand of future technologies. In fact, the relationship between energy output and percentage of Mars’ orbit needing to be covered is linear and as such, if we denote a 50\% coverage of Mars to be the limit of the swarm, this corresponds to a maximum power output of about 155 TW; a consumption of which Earth wouldn’t currently be expected to reach until at least 2100, based on second and third order polynomial fits to historical data from \cite{BP2020} and its previous editions. If our required power output did increase over this estimate, the rate of production of the swarm at this point would be so vast, new satellites could be put into orbits of much larger heights, or even solar orbit, since the reduction in efficiency would be countered by the extreme rate of deployment of the swarm, though numerous energy generation methods would no-doubt be available at this point in our civilisation's development.
        
        Whilst the overall goal of a Dyson Swarm is to provide power for our civilisation, be that on Earth, Mars or beyond, some of the electricity produced will be diverted to the infrastructure on Mars which constructs and launches the swarm itself. To get a better understanding of this process, we can simulate how the energy output of the swarm increases with its size. For this model it is assumed that the initial infrastructure on Mars as well as the first satellite in the swarm have already been built (likely after the first two-year period). The parameters of the model include:
        
            \begin{itemize}
                \item the time to produce a single satellite
                \item the energy required to produce a single satellite, (also accounting for the energy required to continuously increase the capacity of the infrastructure | robots, machinery, etc.)
                \item how often satellites can be launched
                \item the amount of energy being diverted to Mars (`diversion percentage’) and that being sent back to Earth; the latter of which will be denoted the `useful’ output of the swarm.
            \end{itemize}
        
        Whilst the latter variable is varied, the energy to produce a single satellite is set to $10^{12}$ J based on the power consumption of the mass driver, as well as estimates for processes on Earth similar to those which would be carried out on Mars. Due to the changing nature of this value over time, it is difficult to estimate, but $10^{12}$ gives a balanced value for the production across the swarm's lifespan. The time taken to produce one satellite isn’t a significant parameter in the model, even doubling the time has a negligible effect on the overall results, thus a conservative time of four weeks is chosen. The rate at which the swarm is launched is set to one launch every five seconds, per mass launcher, with this particular model having just one mass driver. For each diversion percentage, the model runs until the time elapsed exceeds 125 years but should the active number of satellites surpass 5.51E9 (the number required to make up Earth’s 2019 usage), then no more objects are added to the swarm to allow us to view the data at the values we are interested in. Naturally, the estimate for the energy required to produce one satellite is the most inaccurate. At the same time, it is more likely that the satellites would be continuously produced with the energy available, rather than one at a time, which would increase the rate at which the swarm is built. This adds more uncertainty into the model; however, it is a good approximation and enables us to analyse the parameters.
        
            \begin{figure}
                \centering
                \includegraphics[width=\textwidth]{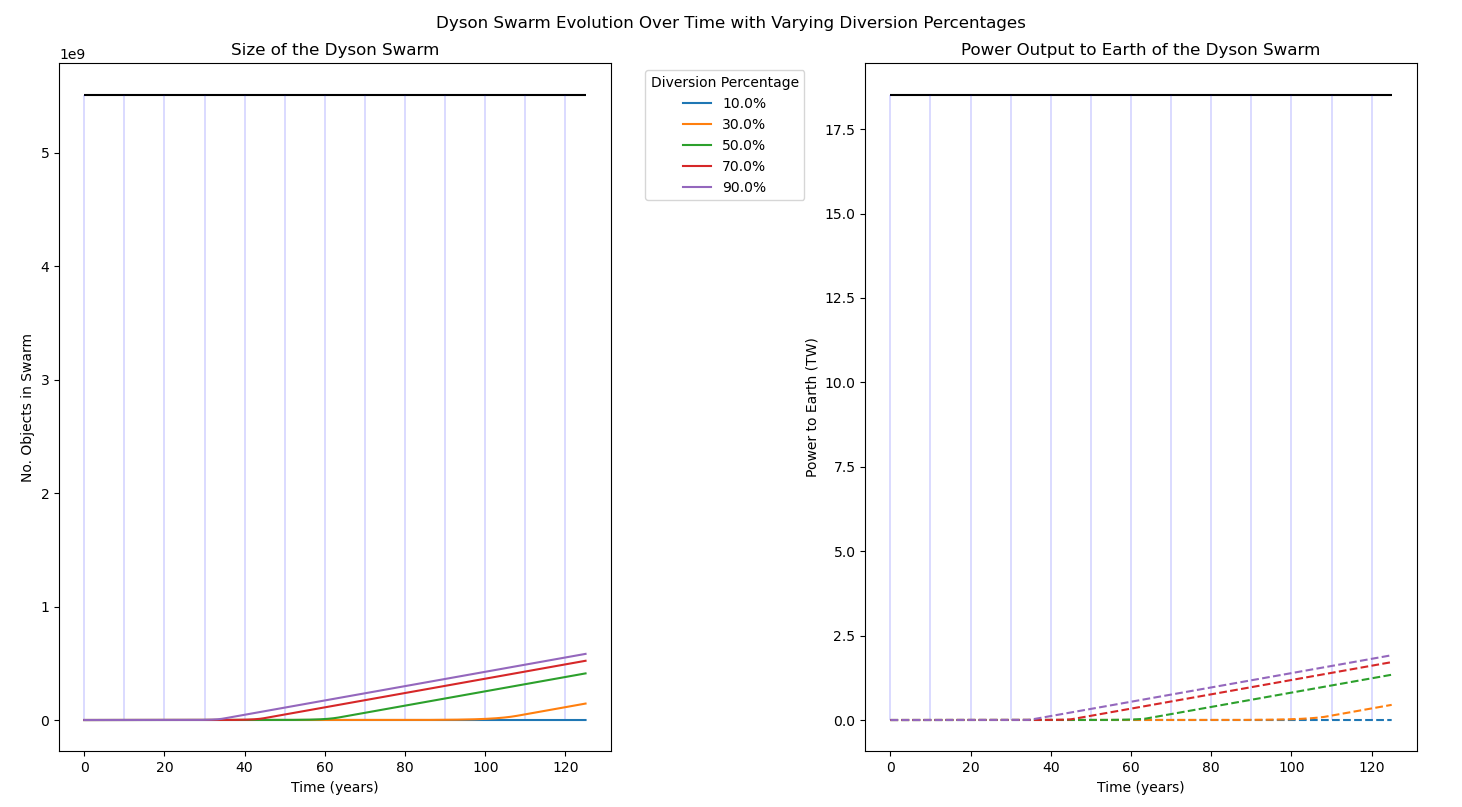}
                \caption{The left graph shows how the number of objects in the swarm which are active and in orbit around Mars increases with time; to the right, the increase in the useful power delivered to Earth. Decades are marked by the vertical light blue lines, the horizontal black lines in both diagrams are the power consumed globally on Earth in 2019, and the size of the swarm required to achieve this. This first model clearly shows that these targets are not met within the 125-year time period.}
                \label{fig:simulation1}
            \end{figure}
        
\begin{figure}
                \centering
                \includegraphics[width=\textwidth]{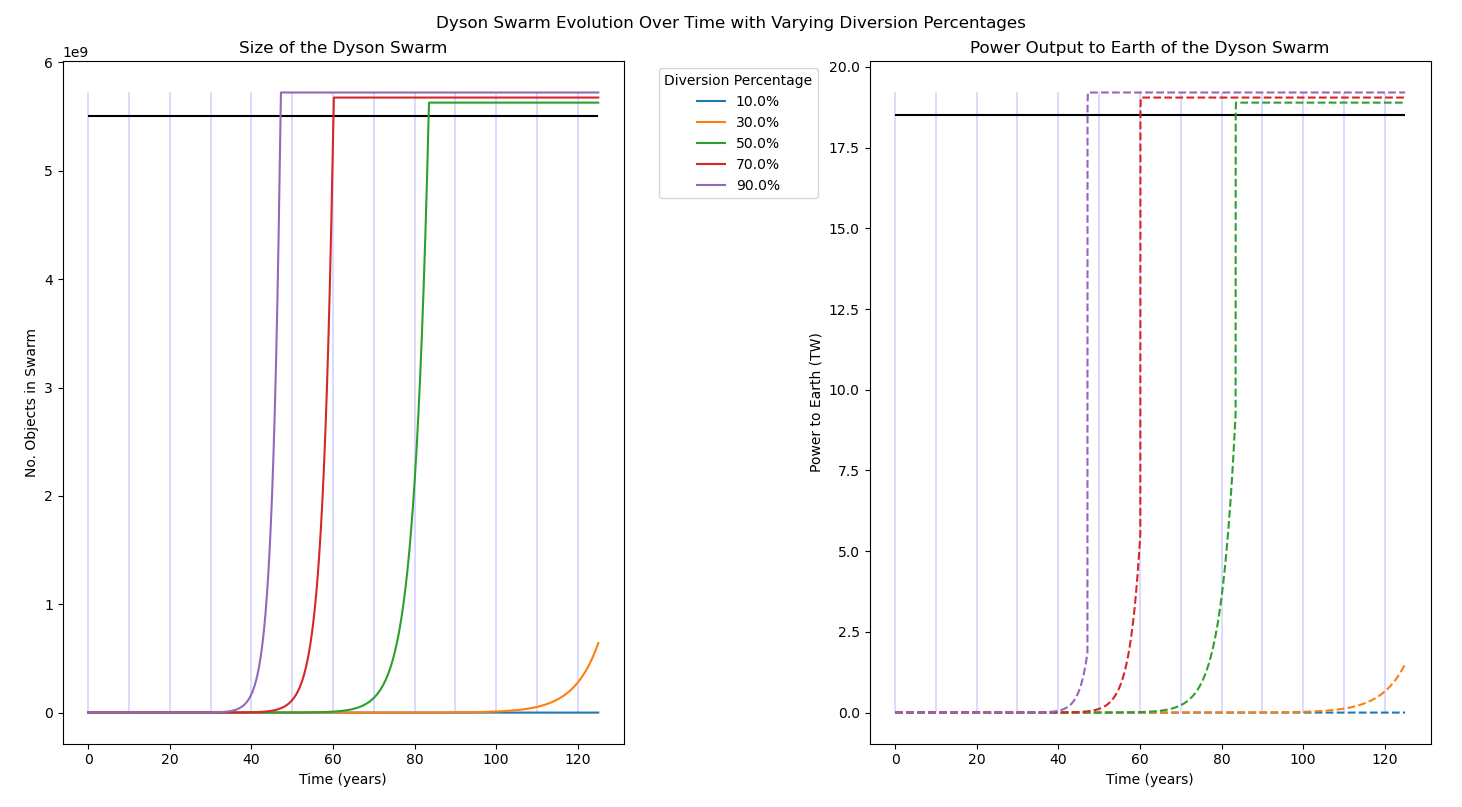}
                \caption{Following the same style as the previous figure, with the continuous addition of mass drivers over time, the power and swarm size for Earth's 2019 consumption are now met for diversion percentages of 50\% and above. Since the model stops the growth of the swarm once these targets have been met, the curves level off which causes this resemblance to the Heaviside Step Function.}
                \label{fig:simulation2}
            \end{figure}

        The results of such a model are shown in Fig.~\ref{fig:simulation1}, alongside a description of its significance; it shows that with the given parameters, even a swarm with a diversion percentage of 90\% would not achieve the capability to supply Earth’s 2019 consumption. The reasoning behind this, however, is firstly that the singular limiting factor is the capacity of the Mars infrastructure at launching the swarm. This is because, whilst the amount of power being diverted to Mars is more than enough to produce extreme numbers of satellites in the swarm, the mass driver(s) are only able to launch at a given rate which maintains linear growth. Secondly, the values of the parameters chosen are very conservative, for example, it is likely that the number of mass drivers would increase with the size of the swarm, which also would allow for a continual increase in the rate at which the satellites can be launched. By implementing this factor into the model such that one mass driver is added per year, Fig.~\ref{fig:simulation2} shows that models with diversion percentages of 90, 70, and 50\% all reach the ideal swarm size for Earth’s 2019 consumption in 47, 60, and 83 years respectively. Exponential growth of the swarm’s size only begins after 115 years at a diversion percent of 30 and doesn’t begin at all within 125 years for 10\%. The exponential growth | which is now present due to the continued addition of more mass drivers | is highly desirable as it easily allows higher energy requirements to be desired in a short period of time. For example, to reach Earth’s estimated energy consumption of 20.35 TW in 2030 from above, corresponds to a swarm size of 6.05E9, which takes no more than a few months to reach than the previous model.
        
\begin{figure}
                \centering
                \includegraphics[width=\textwidth]{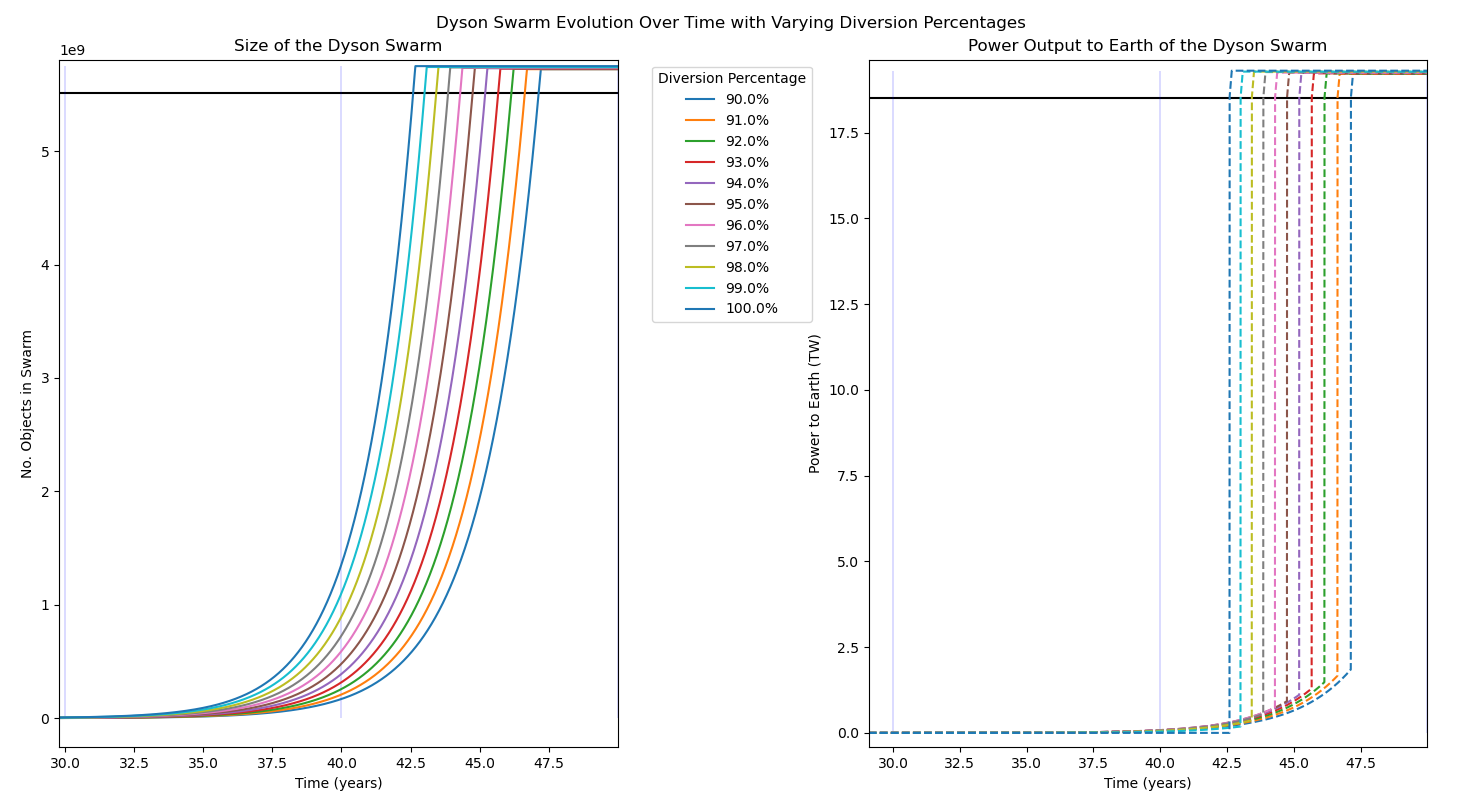}
                \caption{The axes are reduced to between 30 and 50 years such that the differences between diversion percentages can are clearly visible. Whilst a higher percentage means that the swarm reaches the target size a few months earlier than the next integer percentage down, less power is output to Earth during this time. A compromise can be made between time and power as required, and the diversion percentage can change over time too | it does not need to be set and left at a constant value.}
                \label{fig:simulation3}
            \end{figure}

        To determine the optimal amount of energy to keep on Mars and to send to Earth, it is first noted that of the models in Figs.~\ref{fig:simulation1} and \ref{fig:simulation2}, the diversion percentage of 90\% was the best of the five shown. Thus, another model can be generated with percentages ranging from 90 to 100 to see which is the most ideal; this is shown in Fig.~\ref{fig:simulation3}. The higher the diversion percentage, the quicker the swarm can reach its ideal capacity, but the less energy is generated for use on Earth during this time. Any percentage can be chosen to suit required needs; 99\% is chosen as a benchmark such that the swarm can be built in the quickest time whilst also allow small amounts of energy to be produced which may help with the construction, testing and implementation of the WPT technology (on both planets) as the swarm develops.
        
        Due to the exponential growth of both the size of the swarm and the useful power output to Earth, any power consumption in the range of 10 - 100 TW is achieved in around forty-five years. Depending on the amount of power desired to be produced by the swarm, different heights may have to be utilised to prevent overcrowding of Mars' orbits, though by this time, Earth-based renewable energy sources are likely to have progressed massively, as well as the technologies used within the swarm and its construction, enabling its efficiencies to be continually improved. If the first equipment sent to Mars takes two launch windows worth of time (4 years) to set up the initial infrastructure described in section~\ref{section:construc} before production of the swarm can occur, the total time from sending the first spacecraft to Mars and obtaining enough energy to power the planet, is about fifty years. With the first large spacecraft scheduled to land on Mars in 2024, if planning on a Dyson Swarm started today, it’s possible that construction of the swarm could begin in the late 2030s, with an Earth-supporting power output available by 2090.

\section{Conclusion}

        As Dyson himself pointed out, a complete Dyson Sphere is practically mechanically impossible, however, splitting this large structure into numerous smaller components which each contribute to the total output similar to a single sphere, vastly increases the ease of its construction. Similarly, by making the individual satellites of the swarm as simple as possible | a system of reflecting mirrors | the bulk of the technology is placed either on Mars, Earth, or in Earth orbit. This means that as our technologies advance over the roughly fifty-year construction time of the swarm, the capabilities of the whole system can be upgraded continually, increasing its efficiencies and reducing the time to reach desired power outputs, whereas should each satellite need to be upgraded individually, progress would be much slower.

        With the swarm viable, though initially monetarily and materially expensive, only commitment and will is needed to get the wheels turning. With sustainable sources of power becoming an increasingly pressing issue on Earth, as well as our continuously increasing energy demand, a Dyson Swarm could be considered in the near future to combat these vital issues. An increase in renewable energy will aid the energy requirements to get the production of the swarm started before it becomes mostly self-reliant after a decade or so. Embarkation on this task would undoubtedly require a global collaboration, providing countless jobs and expanding research in numerous fields of science. The swarm would not only cater for Earth's power consumptions, but stretch beyond, leading our civilisation through type I towards type II status on the Kardashev scale, allowing us to look further into space exploration and its technologies, for both exploration and habitation as well as understanding more of the mysteries of the universe.

\hrulefill
\begin{multicols}{2}
\section*{Acknowledgements}
    The initial inspiration for this research came many years ago from a Kurzgesagt video titled "How to Build a Dyson Sphere | the Ultimate Megastructure" (see \cite{KURZ}); their videos are extremely thought-provoking and enlightening.

\printbibliography

\end{multicols}
\hrulefill
\end{document}